\documentclass{aa}  

\usepackage{graphicx}
\usepackage{multicol}
\usepackage[normalem]{ulem}
\usepackage[caption = false]{subfig}
\usepackage{subcaption}
\usepackage{hyperref}
\usepackage[svgnames]{xcolor}

\usepackage{txfonts}

\usepackage{natbib}
\bibpunct{(}{)}{;}{a}{}{,} 
\DeclareRobustCommand{\gaia}{\textit{Gaia} }
\DeclareRobustCommand{\teff}{$T_{\mathrm{eff}}$ }
\DeclareRobustCommand{\logg}{$\log g$ }

\newcommand{\kms}{\,km\,s$^{-1}$}

\begin{document}

   \title{Open clusters in the outer disc studied with MEGARA@GTC}

   \subtitle{Auner\,1 and Berkeley\,102}

   \author{J. Carbajo-Hijarrubia\inst{1,2,3}
          \and
          R. Carrera\inst{4}
          \and
          F. Anders\inst{1,2,3}
          \and
          C. Jordi\inst{1},\\
          L. Casamiquela\inst{5}
          \and
          L. Balaguer-N\'u\~nez\inst{1,2,3}
          \and 
          A. Gil de Paz\inst{6,7}
          }

\institute{Institut d'Estudis Espacials de Catalunya (IEEC), Gran Capit\`a, 2-4, 08034 Barcelona, Spain\\
        \email{juan636@fqa.ub.edu}
         \and
        Institut de Ci\`encies del Cosmos (ICCUB), Universitat de Barcelona (UB), Martí i Franquès, 1, 08028 Barcelona, Spain
         \and
        Departament de Física Quàntica i Astrofísica (FQA), Universitat de Barcelona (UB),  Martí i Franquès, 1, 08028 Barcelona, Spain
         \and
        INAF - Osservatorio di Astrofisica e Scienza dello Spazio di Bologna, via Gobetti 93/3, 40129, Bologna, Italy
        \and
        Observatoire de Paris (Meudon) GEPI Batiment 11, Hipparque 5 Place Jules Janssen, 92195, Meudon, France
         \and
        Departamento de Física de la Tierra y Astrofísica, Universidad Complutense de Madrid, Plaza Ciencias, 1, E-28040 Madrid, Spain
        \and
        IPARCOS (Instituto de Física de Partículas y del Cosmos), Facultad de Ciencias Físicas, Ciudad Universitaria, Plaza de las Ciencias, 1, Madrid E-28040, Spain
           }

   \date{}

 
  \abstract
   {Open clusters provide valuable information on stellar nucleosynthesis and the chemical evolution of the Galactic disc, as their age and distances can be measured more precisely than for field stars.}
   {We aim to study the outermost parts of the Milky Way disc using open clusters as tracers. Here we focus on two clusters at Galactocentric distances around 14\,kpc that have never been observed spectroscopically before and are located in largely unexplored regions of the Galaxy.}
   {We use medium-resolution spectra ($R>18\,700$) obtained with the MEGARA integral-field unit (IFU) spectrograph at the 10.4\,m Gran Telescopio de Canarias (GTC) to study red-giant star members of the clusters Auner~1 and Berkeley~102. We determine radial velocities and atmospheric parameters for the member stars, as well as updated ages and distances for these two clusters. Finally, we measure the abundances of six chemical elements: Fe, Ca, Co, Ni, Ba, and Eu.}
   {Both clusters are old, $3.2\pm0.7$\,Ga, distant, $d\sim 8$\,kpc, and moderately affected by interstellar extinction, $A_V\sim 1.3$\,mag, due to their location below the Galactic mid-plane, $Z_{\rm Gal}\sim -0.7$\,kpc. The metallicities of Auner~1, [Fe/H]$=-0.30\pm0.09$, and Berkeley~102, [Fe/H]$=-0.35\pm0.06$, are compatible with the values of other open clusters situated at similar Galactocentric radii, suggesting little azimuthal metallicity variations. The relative abundance ratios, [X/Fe], also behave as expected, perhaps with the exception of [Ca/Fe], which appears slightly enhanced in both clusters, and [Eu/Fe], which is enhanced in Berkeley~102, [Eu/Fe]$=0.64\pm0.05$.}
   {Our results demonstrate the possibility of conducting competitive Galactic archaeology observing with GTC/MEGARA in IFU mode. The two studied objects open a new window into the chemical evolution of the outer Galactic disc. More observations of distant (both in Galactocentric distance and azimuth) open clusters with medium-to-high resolution instruments in 8-10m-class telescopes are needed to firmly establish the abundance trends of the outermost parts of the Galactic disc. }


   \maketitle
%

\section{Introduction}
    \label{intro}
    Open clusters (OCs), found at all ages and throughout the disc, serve as excellent tracers of the overall history of star formation and nucleosynthesis across the Galactic disc lifetime \citep{Friel2013}. This is because, OC ages and distances can be measured more accurately with respect to other tracers like field stars. OCs have been used to trace the spatial distribution of metallicity in the Galactic disc. First studies reported a significant decrease in metallicity with increasing Galactocentric radius, which seems to flatten when reaching a certain distance \citep[e.g.][]{Carrera_Pancino2010} showing a knee at intermediate distances \citep[e.g.][]{magrini2009, Yong2012}. Recent homogeneous studies based on large spectroscopic surveys, GES \citep[\gaia ESO Survey][]{Magrini2022},  APOGEE \citep[Apache Point Observatory Galactic Evolution Experiment][]{myers2022}, GALAH \citep[GALactic Archaeology with HERMES][]{spina2021} and the compilation OCCASO+ \citep[Open Clusters Chemical Abundances from Spanish Observatories][]{Carbajo2024} show that the position of the knee, is located between $R_{\rm GC}\sim$11\,kpc and $R_{\rm GC}\sim$12\,kpc. Nevertheless, of the 187 OCs with more than one star observed spectroscopically at resolution $R$ > 20\,000, only 8 have been studied in the outer part of the Galactic disc ($R_{\rm GC}> 14$\,kpc), which prevents a precise characterization of this change in the gradient. 
    It is difficult to determine abundances for the outer disc OCs, since they are located at large distances, and therefore, the brightness of their stars is faint reaching $G > 15$\~mag. 

\begin{figure*}
  \centering   \includegraphics[width=0.8\textwidth]{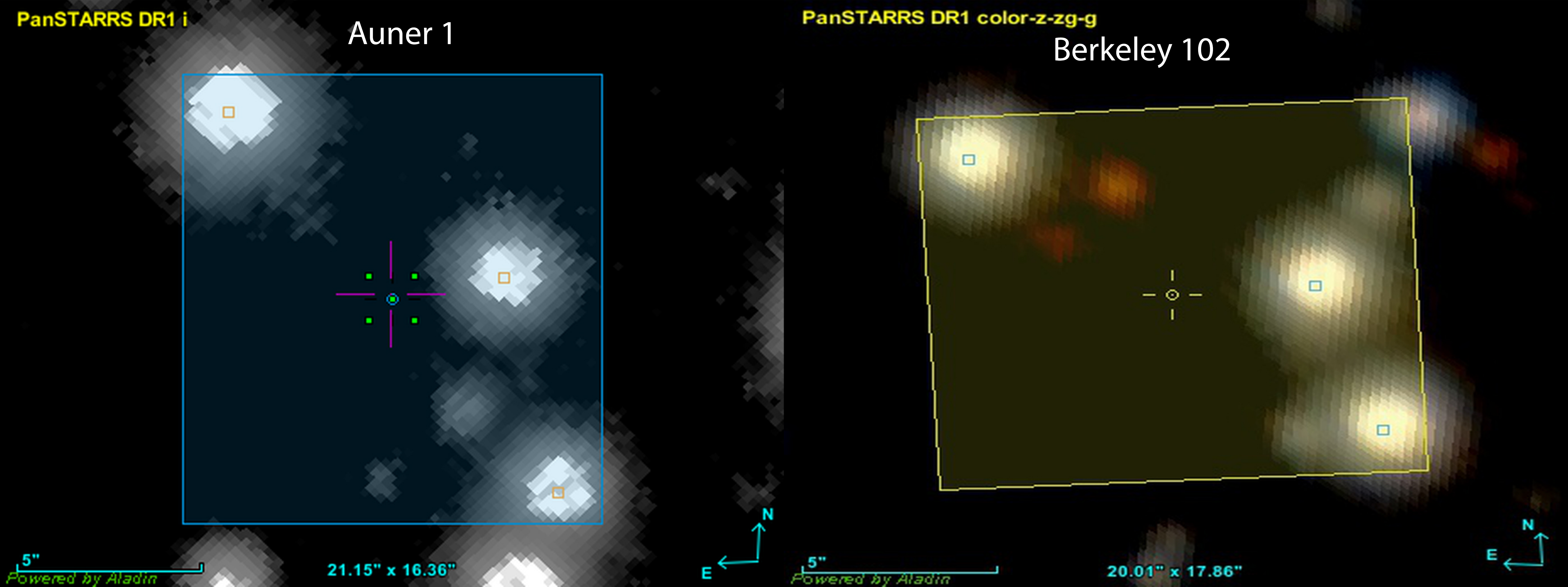}
    \label{fig:subfig2}
  \caption{Spatial distribution in the field of the IFU. We mark with squares the stars observed inside the field of view of MEGARA.}
  \label{fig:Auner1_FOV}
\end{figure*}

    In addition to OCs, several other tracers have been used to study the metallicity gradient, such as \mbox{H\,{\sc ii}} regions \citep[e.g.,][]{HII12011,Arellano2020,mendez_delgado2022}, planetary nebulae \citep[e.g.][]{PN1,Pn2}, Cepheids \citep[e.g.][]{cepheids1,cepheids2,minniti2020,daSilva2022}, low-mass main-sequence stars \citep[e.g.][]{MS1, Anders2017}, massive stars \citep[e.g.][]{Daflon2004,Bragan2019}.
    In almost all cases, a continuously decreasing slope is found, without exhibiting a knee, except in the recent work of \citet{daSilva2023}, which is focused on the study of Cepheids. 
     
    In the present study we obtained observations made from the 10.4\,m Gran Telescopio de Canarias (GTC), and the MEGARA instrument in one of its high-resolution modes at $R\sim$18\,700. We provide, therefore, a study of two OCs in the outer disc, Auner~1 and Berkeley~102. Both clusters are spectroscopically unstudied, and are located at very different Galactic azimuths (see Table \ref{tab:MEGARA_ocs}).  

    The goal of this paper is to study the kinematics of both clusters, their elemental chemical abundances and analyse their metallicities and abundances in the context of the trends derived with other clusters. In this way, we demonstrate the feasibility of conducting competitive Galactic archaeology observations with GTC/MEGARA.  
    The paper is organised as follows. The observations and methodology are described in Sect.~\ref{observations}. We describe the determination of radial velocities and kinematics in Sect.~\ref{MEGARA_RV}, the abundance determination in Sect.~\ref{Sec:MEGARA_AP-ABU}. The Galactic trends are described in Sect.~\ref{OPLUS}. Finally, we discuss our conclusions in Sect.~\ref{sec:conclusions}.  

\section{Observations and methodology}
\label{observations}
     


\label{sec:MEGARA_obs}
\subsection{Instrumentation}
\label{sec:MEGARA_inst}   

MEGARA is an integral field unit (IFU) and multi-object spectrograph (MOS) installed at the 10.4\,m GTC at the Roque de los Muchachos Observatory on the island of La Palma (Spain) \citep{GildePaz2018,Carrasco2018}. MEGARA has 18 diffraction gratings, six of them in low resolution ($R\sim6\,000$), 10 in medium resolution ($R\sim12\,000$) and two in high resolution ($R\sim18\,700$), each covering a different spectral range. These characteristics are the same in both IFU and MOS modes.

The IFU consists of 567 contiguous hexagonal fibres, each with a diagonal of 0.62\,arcsec, encapsulated in a hexagonally-packed rectangular shape, resulting in a field of view of $\sim$12.54\,$\times$\,11.3\,arcsec$^2$. The instrument has eight fibre bundles located on the outer parts of the focal plane, away from the IFU, allowing simultaneous observations of the sky using 56 hexagonal fibres.


We have selected the HR-R Volume Phase Holographic (VPH) grating, since it has a resolution of $R\sim18\,700$ and covers the spectral range $640.56-679.71$\,nm; this allows us to study lines of Fe, Ca, Co, Ni, Ba, and Eu.   
Unfortunately, MEGARA MOS mode was not available during the periods 2022A \& B, so we conducted the study using the IFU mode alone. This limits the number of stars that can be studied per cluster in a single pointing, in comparison with the better-suited capabilities of the MOS mode.

\subsection{Observational strategy}
\label{sec:MEGARA_Estrategia_Observacional}

\captionsetup[subfigure]{labelformat=empty}

\begin{figure*}
  \centering
    \includegraphics[width=0.8\textwidth]{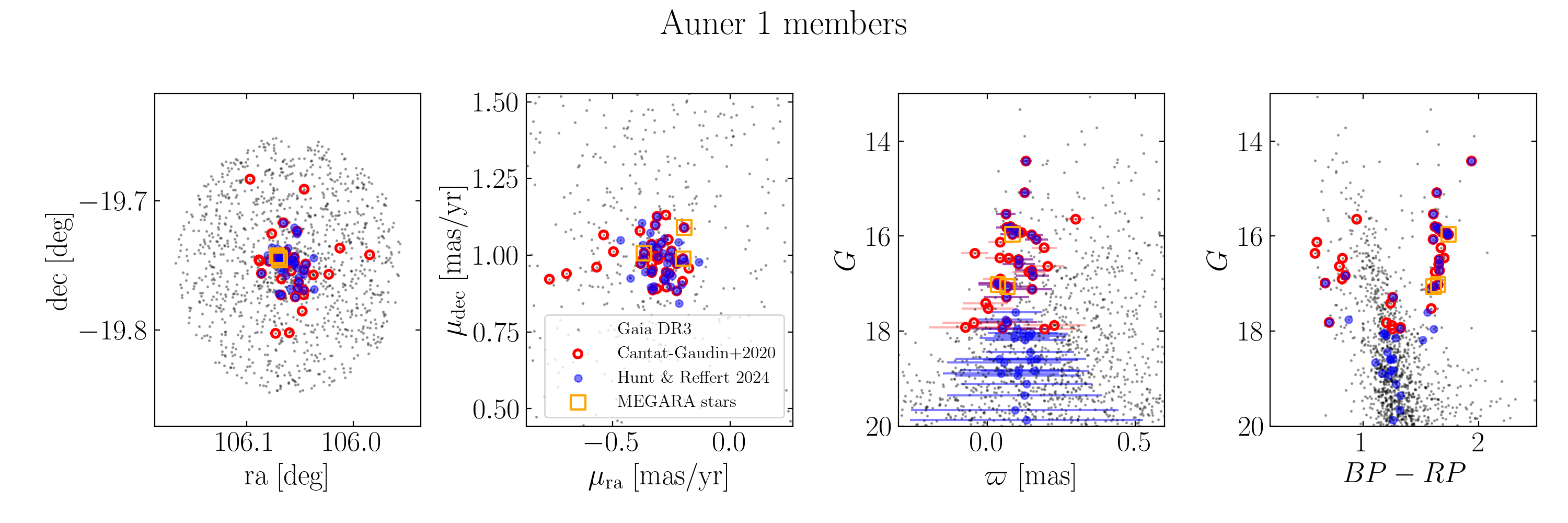}
    \includegraphics[width=0.8\textwidth]{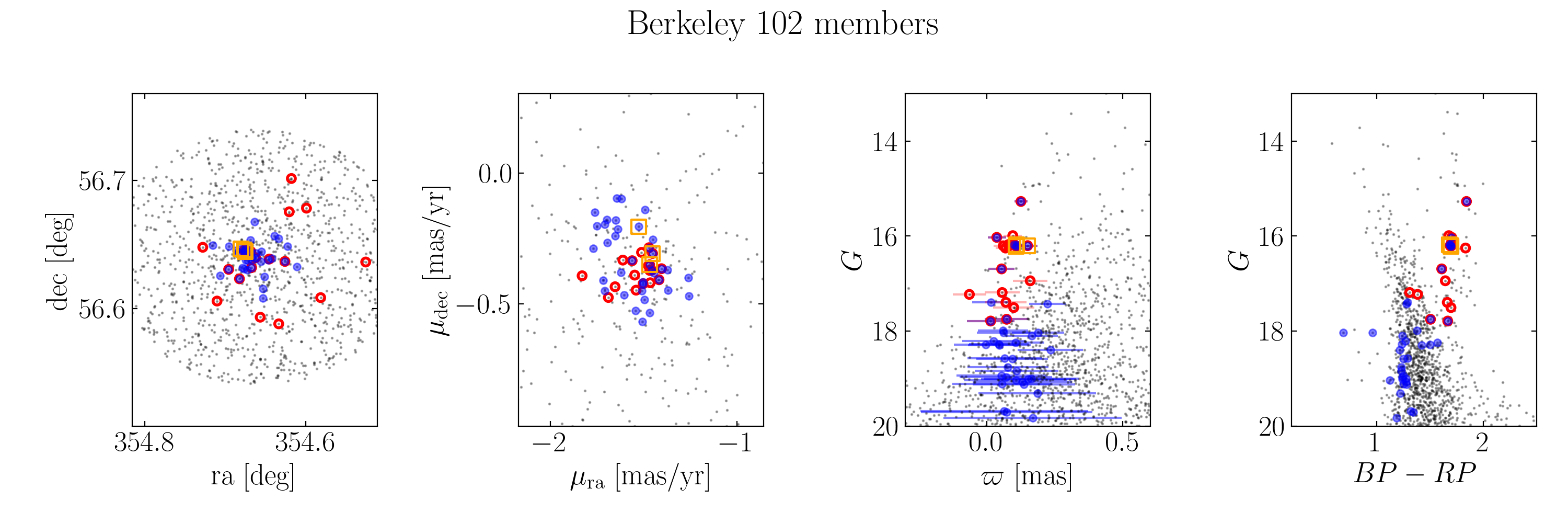}
    \label{fig:subfig2}
  \caption{Sky positions, proper motions, parallax vs. magnitude, and colour-magnitude diagrams (CMD) of Auner~1 (top panels) and Berkeley~102 (bottom panels).
  We show all the stars in the region (black points), the stars considered members of each cluster in \citet{cantat-gaudin2020} (red circles) and in \citet{Hunt2024} (blue circles). We mark the observed stars with orange squares.
  }
  \label{fig:Auner1_pos_CMD}
\end{figure*}

We selected the pointings of the OCs in order to observe three red giant branch (RGB) or red clump (RC) stars in the IFU field of view, using the members of the \citet{cantat-gaudin2020} catalogue\footnote{When the observations were planned, the \citep{Hunt2024} catalogue was not yet available.}. The selection of RC stars is due to the fact that they are in the same evolutionary stage, thus avoiding the variation in abundances between stars caused by stellar evolution or atomic diffusion. The difference in evolutionary stage between RC and RGB stars does not imply a change in the abundances for the elements under study.
RC and RGB stars can be easily identified even in sparsely populated colour-magnitude diagrams (CMDs). They are also brighter than main sequence stars, which allows us to observe them at larger heliocentric distances. As the selected stars are hotter than the brighter giants, their spectra are less crowded with lines, allowing us to determine atmospheric parameters (effective temperature, surface gravity and metallicity) and chemical abundances more precisely.
The observed stars have high astrometric membership probabilities \citep{cantat-gaudin2020,Hunt2024} derived from \gaia EDR3 \citep{Gaia2021}.  



Auner~1 was observed in seven observing blocks (OBs), each with four exposures of 1000\,s (Proposal ID: GTC93-22B). This amounted to a total of 9.57\,h of observation (taking into account the overheads per pointing time). This allowed us to obtain spectra with high enough signal-to-noise ratio (S/N $>$ 40) to derive chemical abundances of three stars with magnitude $G$ between 16 and 17 (Fig. ~\ref{fig:Auner1_pos_CMD} and Table ~\ref{tab:MEGARA_stars}).  
Another star belonging to the cluster was also observed in two of the OBs. The S/N obtained from these two OBs was sufficient to determine $v_{\rm rad}$ but not high enough to measure its chemical abundance.

Berkeley~102 was observed in four OBs, following the same strategy as for Auner~1 (Proposal ID: GTC92-22A). This amounts to a total of 5.5~h of observation, taking into account the pointing time and overheads. In this cluster, we observed three stars with magnitudes $G \sim 16.2$ (Fig.~ \ref{fig:Auner1_pos_CMD}).

\begin{table*}
\setlength{\tabcolsep}{1mm}
\begin{center}
\caption{Details of the observed stars, including information from \gaia DR3, {\tt StarHorse} \gaia EDR3 ($T_{\rm eff}$, \logg), and MEGARA spectra (number of 1000\,s exposures $N_{\rm exp}$, signal-to-noise ratio S/N, $v_{\rm rad}$, $v_{\rm scatter}$, $v_{\rm err}$).}
\begin{tabular}{lcccccccccccccc}
\hline
Star & {\it Gaia} DR3 {\tt source\_id} & $\alpha_{\rm ICRS}$ & $\delta_{\rm ICRS}$  &$G$ & \teff & \logg &$N_{\rm exp}$ & S/N & $v_{\rm rad}$&$v_{\rm scatter}$&$v_{\rm err}$ \\
&  & [deg]& [deg]& [mag] & [K] & [dex]   &  & pix$^{-1}$ & [km\,s$^{-1}$]&[km s$^{-1}$]&[km\,s$^{-1}$]\\

\hline
Auner~1\_1     &2932568798878070528&106.069&-19.746&17.07&4997$^{+96}_{-46}$ &  3.05$^{+0.47}_{-0.08}$&7&41&137.5&1.1&0.7\\
Auner~1\_2     &2932568798878070016&106.070&-19.744&17.02&4930$^{+166}_{-84}$ &   3.1$^{+0.35}_{-0.37}$&6&51&137.9&2.4&0.7\\
Auner~1\_3     &2932568798878068608&106.072&-19.743&15.97&4845$^{+21}_{-8}$ &  1.54$^{+0.01}_{-0.18}$&6&70&136.7&0.5&0.9\\
Auner~1\_4     &2932568798878071552&106.069&-19.746&16.59&4850$^{+122}_{-98}$&3.00$^{+0.21}_{-0.25}$&2&23&136.6&0.7&1.0\\ 
Berkeley~102\_1&1997853073387207296&354.681&56.647 &16.19&4806$^{+172}_{-66}$&2.46$^{+0.05}_{-0.13}$&3&48&-88.9&1.5&0.8\\ 
Berkeley~102\_2&1997853142105532288&354.676&56.646 &16.20&4944$^{+47}_{-181}$&2.65$^{+0.16}_{-0.22}$&4&50&-82.9&0.4&0.9\\ 
Berkeley~102\_3&1997853073386056320&354.676&56.644 &16.21&4923$^{+110}_{-146}$&2.40$^{+0.27}_{-0.03}$&4&50&-88.8&0.4&0.8\\ 
\hline
\end{tabular}
\label{tab:MEGARA_stars}
\end{center}
\end{table*}


\subsection{Data reduction}
\label{sec:MEGARA_reduccion_datos}  
The raw IFU spectra of both OCs were reduced using MEGARA data reduction pipeline \citep[MDRP,][]{Pascual2022}. We started by removing the level of \textit{bias} present in the image. To remove this value from the counts in the images, we use \textit{bias} images taken during the corresponding observing night and the task {\tt MegaraBiasImage}.         
Faulty pixels are automatically masked in this step with the file {\tt master\_bpm.fits}, available at MDRP. The spectra are then processed using the {\tt Trace} and {\tt ModelMap} tasks, tracing the fibre spectra through the halogen lamp images. MDRP was used to perform wavelength calibration of the spectra using ThNe calibration lamp images. Using MDRP we performed a fiber-flat correction, using the same halogen lamp images.
We combine the calibrated exposures of each OB using the median, and rejecting cosmic rays in the process. In addition, MDRP subtracts the mean spectrum of the sky, determined by spectra of the dedicated fibres, to generate the fully reduced spectra. For each OB we have selected all the spaxels on which the signal from the corresponding star was clearly seen above the noise. For this purpose, we have used the MEGARA QLA code \citep{Gomez-Alvarez2018}. This implied that, for the range of seeing of our observations (between 0.7 and 1.6), we had to extract two rings of hexagonally-shaped and packed spaxels around the brightest spaxel on which the star is centred, or 1+6+12=19 spaxels per star.


The following processing steps are performed using the code \texttt{iSpec} \citep{blancocuaresma2014}. We combine spectra belonging to the same star observed in different OBs. 
We determine the continuum level of the real spectra by comparison with a synthetic reference spectrum generated with \texttt{iSpec} and then normalise. We use the radiative transfer code and line list described in Sect.~\ref{Sec:MEGARA_AP-ABU}. The reference spectra are computed using the \teff and \logg taken from the \texttt{StarHorse} catalogue \citep[][Table~\ref{tab:MEGARA_stars}]{Anders2022} and the metallicity is fixed for all stars to an estimation for the Galactocentric radius of the clusters ([M/H]=-0.2\,dex).
This process is done in conjunction with the determination of the $v_{\rm rad}$ of the star, as detailed in Sect. ~\ref{MEGARA_RV}.
\texttt{StarHorse} derive the stellar properties with the photo-astrometric information of {\it Gaia} EDR3 combined with the photometric catalogues of Pan-STARRS1 \citep{Chambers2016}, SkyMapper \citep{Onken2019}, 2MASS \citep{Cutri2003}, and AllWISE \citep{Cutri2013}. 



\section{Cluster parameters and kinematics}
\label{MEGARA_RV}

In this section, we use {\it Gaia} DR3 data together with our GTC/MEGARA observations, to determine precise 3D positions and kinematics of the two distant clusters.

\subsection{Redetermination of distances and ages}
\label{Sec:MEGARA_dist-age}

We begin by redetermining the clusters' astrophysical parameters in order to obtain more accurate distances and ages, following the method used in \citet{Anders2022b}.
Although the list of members of \citet{Hunt2024} is based on {\it Gaia} DR3 and the list by \citet{cantat-gaudin2020} is based on {\it Gaia} DR2, the stars that we observed belong to both lists. Thus, both works consider our stars as members of the clusters with high probability (Fig.~\ref{fig:Auner1_pos_CMD}).
Using the \citet{Hunt2024} members, we first redetermine the zero-point corrected cluster parallaxes. We calculate the parallax zeropoint for all stars in the field (following \citealt{Lindegren2021}), apply this correction to the member stars, and then determine the corrected mean parallax (and its uncertainty) for the two clusters as $\varpi_{\rm Auner 1}^{\rm corr}=0.134\pm0.006$\,mas and $\varpi_{\rm Be 102}^{\rm corr}=0.121\pm0.010$\,mas. 
Besides, we calculate Bayesian distances to the clusters, assuming negligible parallax correlations, a zero-point uncertainty of 0.005\,mas, and a negligible cluster extent (compared to the distance), finding $d_{\rm Auner 1}=7.46^{+0.66}_{-0.57}$\,kpc and $d_{\rm Be 102}=8.26^{+1.16}_{-0.91}$\,kpc.

\begin{figure*}
  \centering
    \includegraphics[width=0.49\textwidth]{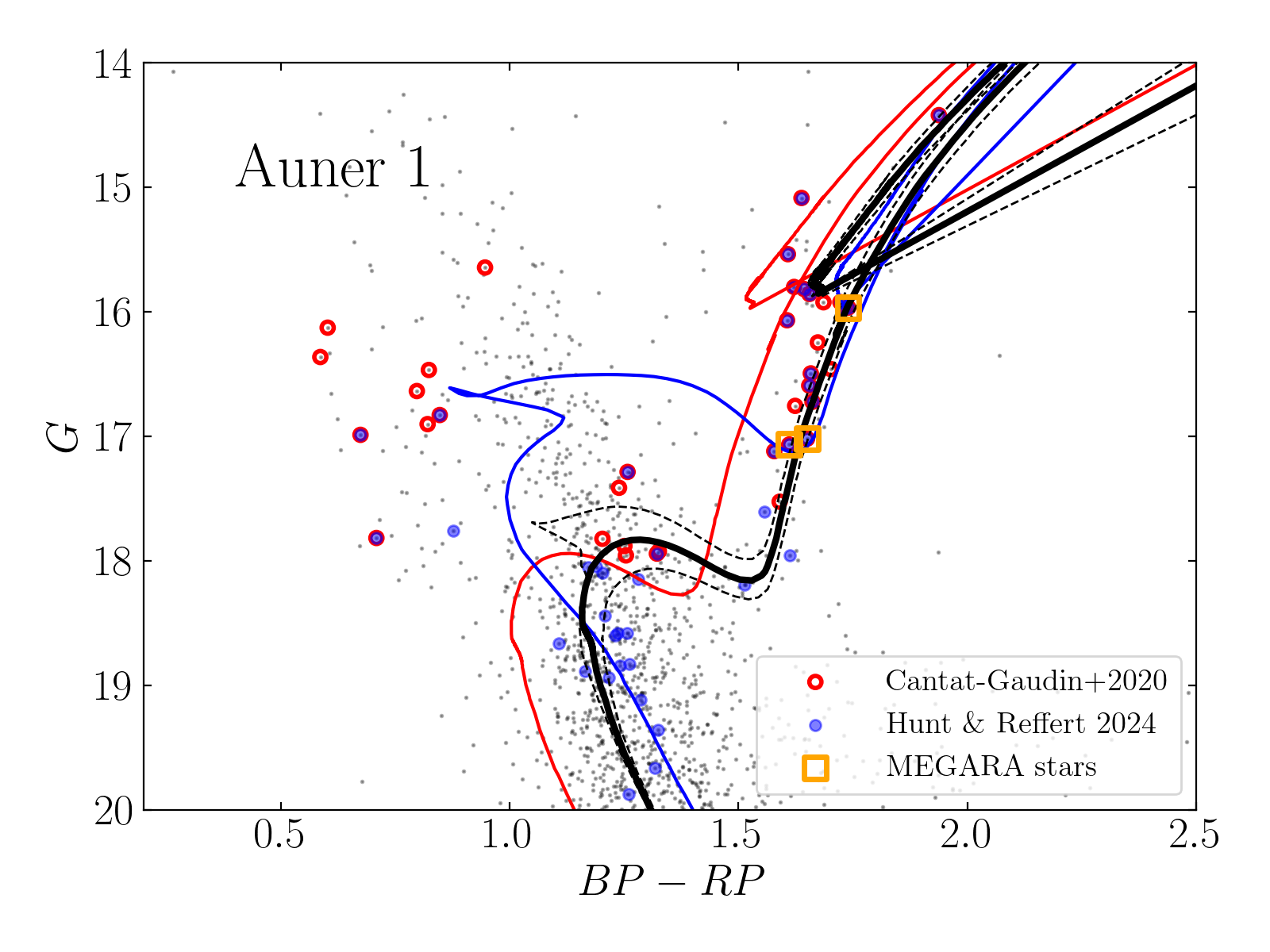}
    \includegraphics[width=0.49\textwidth]{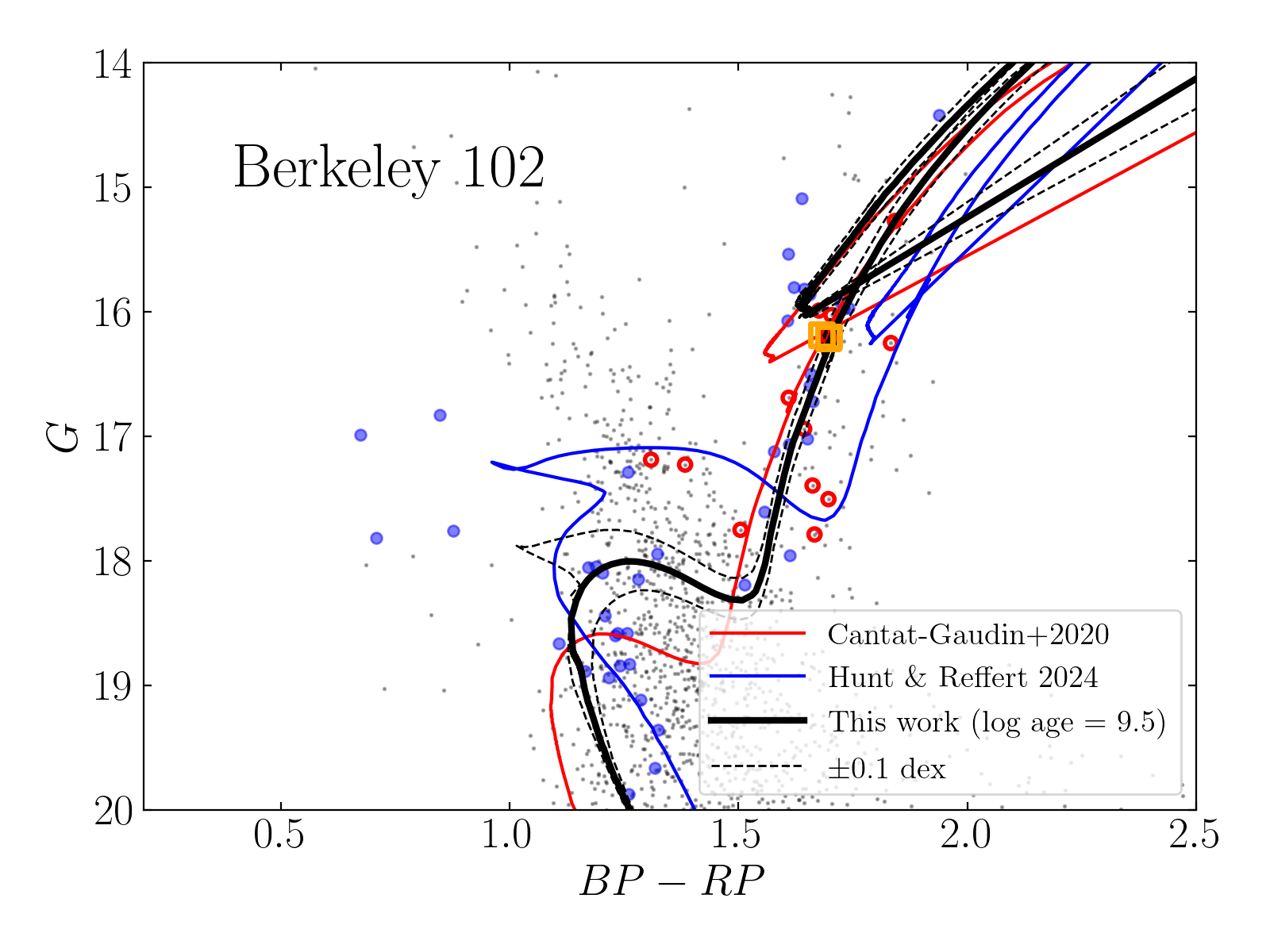}
    \label{fig:subfig2}
  \caption{CMD of Auner~1 (right) and Berkeley~102 (left). Stars in the field are marked as grey dots, those considered members by \citet{cantat-gaudin2020} as red circles, those by \citet{Hunt2024} as blue circles. We also mark the observed stars with orange squares (those in Berkeley~102 are on top of each other). We superimpose the isochrone fitting of \citet{cantat-gaudin2020} (red), \citet{Hunt2024} (blue), and ours (black).}
  \label{fig:CMD}
\end{figure*}

We redetermine the interstellar extinction towards the two clusters using the results from \texttt{StarHorse} for \textit{Gaia} EDR3 stars \citep{Anders2022}, crossmatched with the \citet{Hunt2024} membership catalogue. We derive extinctions of $A_V^{\rm Auner 1} = 1.35\pm 0.03$\,mag and $A_V^{\rm Be 102} = 1.32\pm 0.12$\,mag, respectively. With extinction and distance precisely determined, and with metallicity derived from MEGARA spectra (see Sect.~\ref{Sec:MEGARA_AP-ABU}), we revisit the cluster CMD and adjust PARSEC 2.0 isochrones \citep{Nguyen2022} to determine the best-fitting ages of the two objects (see Fig.~\ref{fig:CMD}). In both cases, isochrone with $\log{ age} = 9.5$ results in the best fit of the blue edge of the main sequence, main-sequence turn-off and red clump. We estimate the uncertainty in these visual fits around 0.1\,dex in log age (see dashed lines in Fig. \ref{fig:CMD}), so we obtain an age of $3.2\pm0.7$\,Ga for both clusters.
The physical parameters for Auner~1 and Berkeley~102 are summarised in Table \ref{tab:MEGARA_ocs}.


\subsection{Measurement of the radial velocity}

MEGARA is not thermally stabilised, so it can have small variations in the wavelength calibration, which affects the radial velocity determination. To characterise these variations, we have used the sky spectra obtained in each OB. Based on the sky emission line list of \citet{Hanuschik2003}, from UVES instrument spectra, we prepared a synthetic sky spectrum with the resolution and spectral range of the MEGARA HR-R VPH grating. With the 24 lines available, we calculated the $v_{\rm rad}$ differences of each OB with respect to the reference spectrum by cross-correlation using \texttt{iSpec}. We find that the typical differences between observations are of the order of 2.0$\pm$0.7\,km\,s$^{-1}$ reaching a maximum of 6.5$\pm$0.7\,km\,s$^{-1}$. Each OB is corrected for the differences identified. We have verified, comparing the dispersion of results using and not using this procedure, that it improves RV accuracy by a factor of ten.

Once the wavelength calibration is refined, we apply the heliocentric correction to the spectra. We measure the velocity difference between the OBs by cross-correlation, using as reference the observation with the highest S/N. We correct the differences in $v_{\rm rad}$ and combine the spectra. In this step we also measure the dispersion in the measurements of $v_{\rm rad}$ of each OB ($v_{\rm scatter}$) taking into account the low statistics correction factor in $\sigma$ by applying the equation~5 of \citet{Roesslein2007}. We also derive the uncertainty in the determination of $v_{\rm rad}$ by cross-correlation with the reference spectrum ($v_{\rm err}$). 
We determine the radial velocity of the combined spectrum with \texttt{iSpec} by cross-correlation with the same reference spectra used in the normalisation process. 
We have verified that, by using a reference spectrum with the metallicity measured in Sect.~\ref{Sec:MEGARA_AP-ABU}, we obtain the same value of $v_{\rm rad}$.
The results are presented in Table~\ref{tab:MEGARA_stars}.

In the cases of stars Berkeley\,102\_1, Auner\,1\_1 and Auner\,1\_2, the value of $v_{\rm scatter}$ is higher than that of the other stars. The S/N values are not low enough to explain these differences. The $v_{\rm scatter}$ values for those stars could be representative of the uncertainties, although it could be also a sign of binarity.
Because no spectral lines of companion stars were detected in the spectra, we also use these stars to determine chemical abundances in Sect.~\ref{Sec:MEGARA_AP-ABU}. 
Without considering the star with the highest $v_{\rm scatter}$, the mean value of $v_{\rm scatter}$ is 770\,$\rm ms^{-1}$.
Berkeley\,102\_2 has a considerably different value of $v_{\rm rad}$ than the other two observed stars. The membership of this star should be reviewed once the $v_{\rm rad}$ of more stars in the cluster is measured. We find the $v_{\rm rad}$ of the other observed stars to be compatible with each other, confirming that they are member stars of the respective clusters.

\subsection{Membership and mean radial velocity of clusters}
\label{sec:members}

To determine the radial velocity of the cluster, $v_{\rm rad, OC}$, we use a weighted average, following the same procedure as in \citet{Soubiran2018}. The mean $v_{\rm rad, OC}$ is obtained using:

\begin{equation}\label{eq:meanrv}
v_{\rm rad,OC} = \frac{\sum_i v_{{\rm rad,}i} \times w_i}{\sum_i w_i}
,\end{equation}
where $v_{{\rm rad,}i}$ is the radial velocity of each star in the cluster and the weight $w_i$ is defined as $w_i=1/(v_{{\rm scatter,}i})^2$. 
    
Similarly, the internal velocity dispersion is derived as:

\begin{equation}\label{eq:s_rv}
\sigma_{v_{\rm rad,OC}}=\sqrt{\frac{\sum_{i} w_{i}}{(\sum_{i} w_{i})^2+\sum_{i} w_i^2}\times\sum_i w_i\times(v_{\mathrm{rad},i}-v_{\rm rad,OC})^2}
\end{equation}

\vspace{1.5cm}

Finally, the uncertainty in the mean $v_{\rm rad, OC}$, $e_{v_{\rm rad, OC}}$ defined as the maximum of the standard error, $\frac{\sigma_{v_{\rm rad, OC}}}{\sqrt{N}}$, and $\frac{I}{\sqrt{N}}$ \citep[][]{Jasniewicz1988}, where $N$ is the number of member stars and $I$ is the internal error of $v_{\rm rad,OC}$ defined as:
    
\begin{equation}\label{eq:interror}
I = \frac{\sum_i w_i \times v_{\mathrm{scatter}, i}}{\sum_i w_i}
\end{equation}

The values obtained are listed in Table~\ref{tab:MEGARA_ocs}. This is the first time that $v_{\rm rad}$ has been determined for these clusters.

To investigate the kinematics of the clusters in the context of the Galactic disc, we study the radial velocity of the OCs along with the proper motions, distances, and ages given in Table \ref{tab:MEGARA_ocs}. We use mean proper motions from \citet{Hunt2024} based on \textit{Gaia} DR3 data. 

\subsection{Radial velocity with respect to GSR and RSR}
\label{sec:RGRS}

We calculate the line-of-sight velocity with respect to the Galactocentric Standard of Rest (GSR) and with respect to the Regional Standard of Rest (RSR), using the following equations:

\begin{align}
\begin{split}
v_\text{GSR}=  &v_\text{rad,OC}+U_{\odot}\cos l \cos b \\
 &+ \left( \Theta_{\text{0}}+V_{\odot} \right) \sin l \cos b + W_{\odot} \sin b,
\end{split}
\end{align}
\begin{align}
v_\text{RSR}=v_\text{GSR}-  \Theta_R \frac{R_0}{R_{\rm GC}} \sin l \cos b,
\end{align}

\noindent where $v_\text{rad,OC}$ is the mean heliocentric radial velocity of the ccluster, derived in the previous Sect.~\ref{sec:members}. ($U_{\odot}$, $V_{\odot}$, $W_{\odot}$)
are the components of the Sun's motion with respect to the Local Standard of Rest (LSR), and $\Theta_0$ and $\Theta_R$ are the circular velocities at Galactocentric distances from the Sun ($R_0$) and the cluster ($R_{\rm GC}$), respectively.
For the Sun, we adopt $(U_{\odot},V_{\odot}, W_{\odot})=$ (11.1, 12.24, 7.25)\,km\,s$^{-1}$, as obtained by \citet{Schonrich} and $R_0$\,=\,8.34\,kpc derived in \citet{Reid2014}. As the circular velocity around the Galactic centre, we adopt $\Theta_{\text{0}}$= 240\kms~ from \citet{Reid2014}, and $\Theta_{\text{R}}$ is calculated according to the Galactic potential described below, in Sect.~\ref{sec:orbita}. 

The results are listed in Table~\ref{tab:MEGARA_ocs}. The uncertainties were estimated with 100\,000 random draws from uncorrelated Gaussian uncertainties, taking into account errors in the radial velocities and cluster distances. The two clusters show values of $v_\text{RSR}$ typical of clusters with thin-disc kinematics \citep{Carrera2021}. 

\begin{figure}
\centering
\includegraphics[width=\columnwidth]{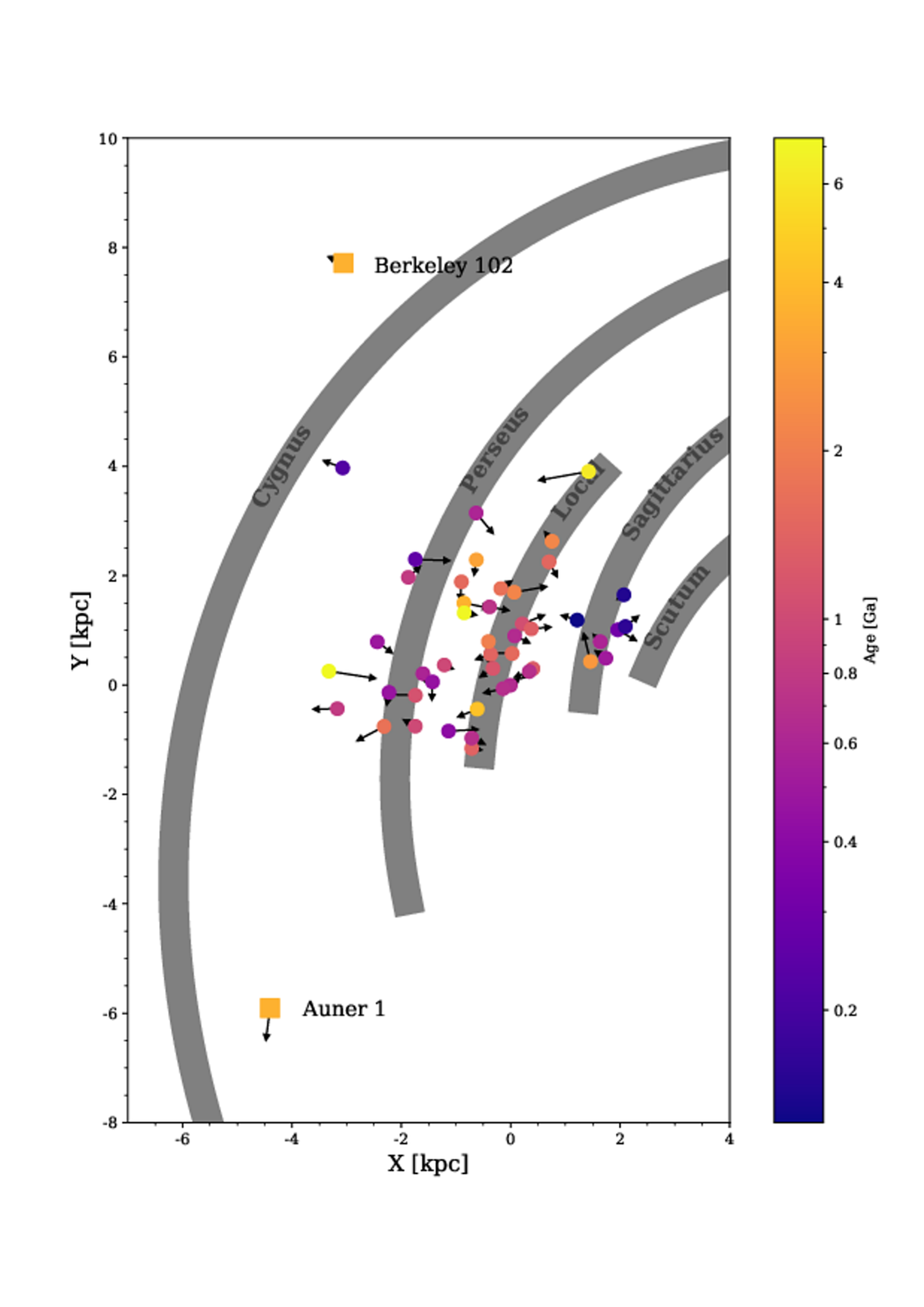} 
\caption{Projection on the Galactic plane of the position and velocity with respect to the Regional Standard of Rest of our two clusters (squares) and \citet{Carrera2021} (circles), coloured by age. We also depict the Spiral Arms determined by \citet{reid2019}.}
\label{fig:ux_vy}
\end{figure}

We derive full spatial velocities with respect to the GSR $(V_\text{R}, V_{\phi}, V_\text{z})$ and RSR $(U_\text{s},V_\text{s},W_\text{s})$ being $U_\text{s}=V_\text{R}$, $V_\text{s}=V_\phi-\Theta_R$ and
$W_\text{s}=V_\text{z}$. These values are also included in Table~\ref{tab:MEGARA_ocs}.  Figure~\ref{fig:ux_vy} shows the projection onto the Galactic plane of the position and velocity with respect to the RSR of the clusters in our sample, in addition to those studied in \citet{Carrera2021} included for comparison. We also represent the Spiral Arms determined by \citet{reid2019}.  
Auner~1 and Berkeley~102, have peculiar velocities compatible with the kinematics of the thin disc, as we have also seen when studying their $v_\text{RSR}$. 

\begin{table*}[h!]
\setlength{\tabcolsep}{0.8mm}
\caption{Properties of the two studied clusters. First rows: Main cluster parameters derived from {\it Gaia} DR3 (see Sect.~\ref{Sec:MEGARA_dist-age} for details). Middle rows: Kinematics (Sects.~\ref{sec:members} through \ref{sec:orbita}). Bottom rows: Chemical abundances derived from MEGARA spectra (Sect.~\ref{Sec:MEGARA_AP-ABU}).}
\begin{tabular}{lccccccccccccccc}
Cluster        & $\alpha_{\rm ICRS}$ & $\delta_{\rm ICRS}$  & Age &  Distance & \emph{X} & \emph{Y} & \emph{Z} & $R_{\rm GC}$  & $A_V$\\
        & [deg]    & [deg] &  [Ga] &  [kpc] & [kpc] & [kpc] & [kpc] & [kpc] & [mag] \\
\hline
Auner 1&106.06&$-19.74$8&3.2$\pm$0.7&7.46$^{+0.66}_{-0.57}$&$-12.7\pm0.4$&$-5.9\pm0.5$&$-0.78\pm0.06$&$14.1\pm0.5$&$1.35\pm0.03$\\
Berkeley 102&354.66&56.637&3.2$\pm$0.7&$8.26^{+1.16}_{-0.91}$&$-11.40\pm0.05$&$7.7\pm1.1$&$-0.68\pm0.10$&$13.8\pm1.0$&$1.32\pm0.12$\\

\hline
\end{tabular}
\vspace{0.5cm}
\newline
\newline
\begin{tabular}{lcccccccccccc}

  & $N_{\rm RV}$& $v_{\rm rad,OC}$ & $\sigma_{v_{\rm rad,OC}}$ & $e_{v_{\rm rad,OC}}$ &  $v_\text{GSR}$  &   $v_\text{RSR}$ &   $U_\text{s}$      &   $V_\text{s}$ & $W_\text{s}$  & $V_{\phi}$\\
      & & [\kms] & [\kms] & [\kms] & [\kms] & [\kms] & [\kms] & [\kms] &[\kms] & [\kms]\\
\hline  
  Auner\,1 & 4&137.17 & 0.66 & 0.33 & -68.0$\pm$0.7  &25.3$\pm$2.7   &  -25.5$\pm$2.8  & -38.5$\pm$2.8 &  0.8$\pm$5.9    & 182.9$\pm$2.9\\
  Berkeley\,102 &3& -86.85 & 3.45 & 1.99 & 139.6$\pm$3.6  &30.3$\pm$4.2   &  -22.8$\pm$4.1  & -4.2$\pm$3.1  &  20.4$\pm$4.1   & 217.1$\pm$3.4\\ 
\hline
\end{tabular}
\vspace{0.5cm}
\newline
\newline
\begin{tabular}{lcccccccc}
  & $N_{\rm ABU}$ &[Fe/H] &[Ca/Fe] & [Ti/Fe] &[Co/Fe] &[Ni/Fe] &[Ba/Fe] &[Eu/Fe]\\ 
        & &[dex] &[dex] &[dex] &[dex] &[dex] &[dex] &[dex]\\ 
\hline
Auner 1& 3& -0.30 $\pm$ 0.10 &   0.17 $\pm$ 0.18 &  0.20$\pm$0.26\tablefootmark{a} &  0.05 $\pm$ 0.17\tablefootmark{a} &  0.05 $\pm$ 0.08 &  0.22 $\pm$ 0.10&   -- \\
Berkeley 102 & 3 & -0.35 $\pm$ 0.07 &  0.19 $\pm$ 0.15 & 0.12$\pm$0.09 &-0.02 $\pm$ 0.10 &    0.01 $\pm$ 0.05 &  0.33 $\pm$ 0.14 &  0.64 $\pm$ 0.05\tablefootmark{a} \\
        \hline
\end{tabular}
 \tablefoot{\tablefoottext{a}{Base only on two stars.}}
\label{tab:MEGARA_ocs}
\end{table*}

\subsection{Orbital analysis}
\label{sec:orbita}
To complete our analysis, we integrate the orbits of Auner~1 and Berkeley~102. Due to the uncertainty in the characterisation of the real Galactic potential, we consider four models proposed in the literature. 
The one proposed by \citet{Bovy2015}, denoted as MW2014, is an axisymmetric potential composed of a spherical bulge, a Miyamoto-Nagai disc, and a halo with a Navarro-Frenk-White profile \citep{nfw97}. The second and third models are based on the previous one, but with the addition of two non-axisymmetric components. For the second model we add a bar characterised by a Ferrers' potential \citep{Ferrers1877} with $n=2$, the semi-major, semi-medium and semi-minor axes are set to 3\,kpc, 0.35\,kpc and 0.2375\,kpc, respectively; the mass of the bar is $10^{10}$\,M$_{\odot}$ \citep{Romero-Gomez2015} and a constant velocity pattern set at $\Omega = 42$\,km s$^{-1}$\,kpc$^{-1}$ \citep{Bovy2019} is used, which places the corrotation at $R_{\rm GC} =5.6$\,kpc and the outer Lindblad resonance at $R_{\rm GC} = 9$\,kpc.
The angular orientation of the bar with respect to the Sun-Galactic centre line is $20^{\rm o}$ \citep[see][and references therein]{Romero-Gomez2011}. 
The third model adds spiral arms sinusoidal spiral of \citet{Cox-Gomez2002} to the axisymmetric potential. We model two spiral arms with an amplitude of 0.1 and a velocity pattern of $\Omega = 21$~km s$^{-1}$\,kpc$^{-1}$ \citep[e.g.,][]{Antoja2011}, which puts the corrotation at $R_{\rm GC}=10.6$\,kpc. 
The fourth model is the sum of the axisymmetric potential with the bar and arms described above.      
Using the package \texttt{Python} \texttt{galpy} \citep{Bovy2015}, we have integrated the orbit backwards in time over the cluster age, with a step of 2\,Ma. The components of the Sun's motion with respect to the LSR, the Galactocentric distance from the Sun, the circular velocity at this distance, and the distances and ages of the OCs are as described in Sect.~\ref{sec:RGRS}.  

\begin{figure}
\centering
\includegraphics[width=\columnwidth]{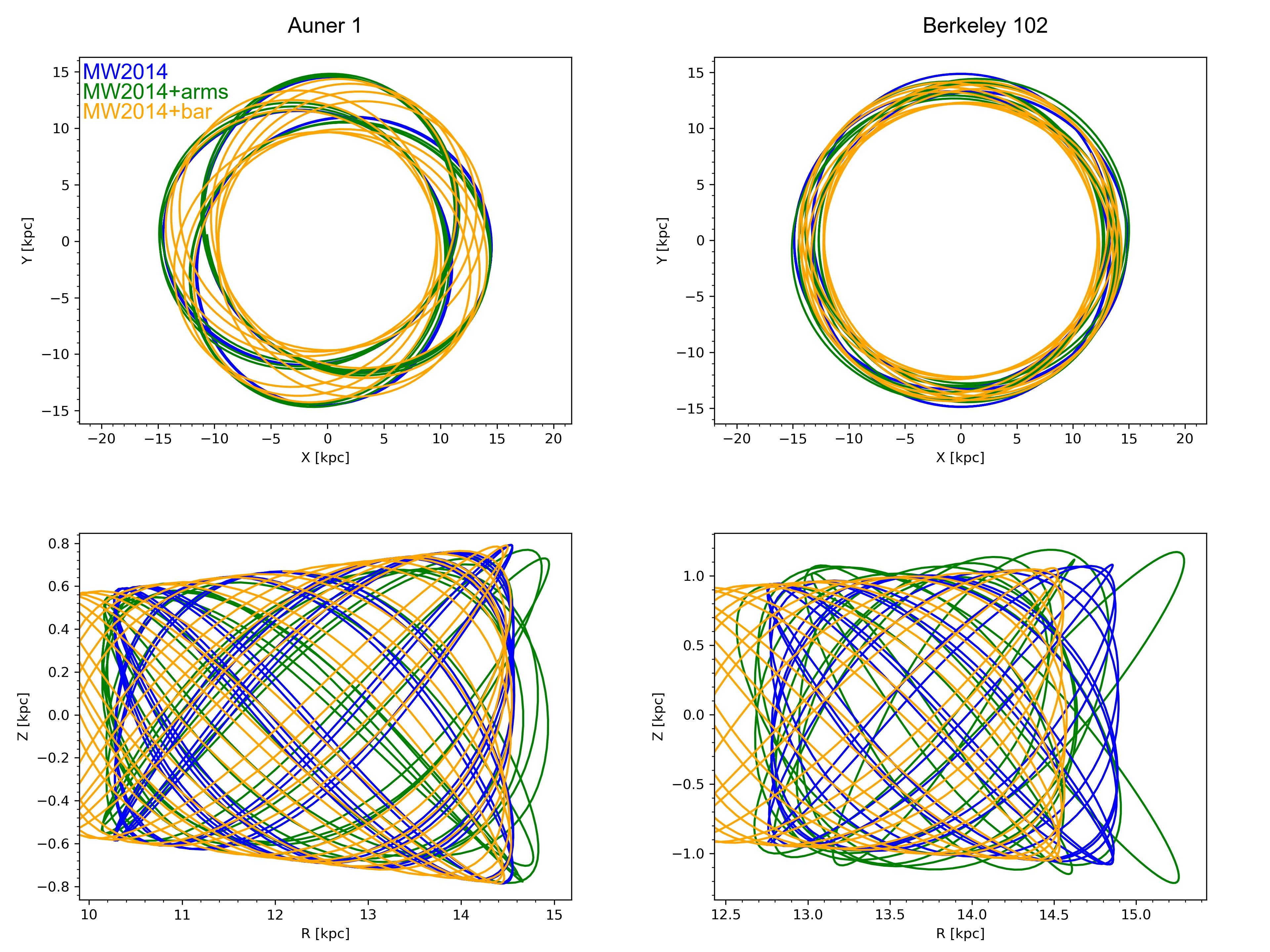}
\caption{Orbits of Auner~1 (left) and Berkeley~102 (right), for potentials MW2014 (blue), MW2014 + arms (green), MW2014 + bar (yellow).}
\label{fig:OC_MEGARA_orbitas}
\end{figure}

It is important to note that, for old clusters, orbital calculations have large uncertainties, which increase the further back in time the orbit is determined. This is because we do know neither the temporal evolution of the potential, nor the interaction of the clusters with structures in the disc such as molecular clouds.
To calculate the uncertainties of the orbital parameters, we have analysed each orbit for the possible values of each of the input parameters in play, taking into account their uncertainties. We used 1000 realisations of Monte Carlo sampling values and integrated the orbits. We show the most probable orbits of both clusters assuming the potentials MW2014, MW2014+bar and MW2014+arms in the X-Y and $R_{\rm GC}$-Z planes (Fig.~\ref{fig:OC_MEGARA_orbitas}), and we list the orbital parameters computed with all the potential in Table~\ref{tab:Orbitas}.   

We explore the parameters that can be fitted to the potential, one of which is the amplitude of the arm potential \citep{Cox-Gomez2002}. This factor indicates the importance of the arm potential with respect to the axisymmetric potential. The value used in the \citet{Cox-Gomez2002} standard model is 40\,\%. We find that this value yields large displacements in $Z$ for Auner~1 and Berkeley~102 and high eccentricities, around 0.3.
Although an arm amplitude of 40\,\% may fit well other galaxies, assuming an arm amplitude of 10\,\% agrees better with the literature values for the Milky Way and produces orbits more similar to those calculated with the other potentials. 
The orbits computed with the different potentials (including those amusing spiral arm potential with a 10\,\% amplitude) are the orbits expected for objects in the thin disc \citep[][see Fig.~\ref{fig:OC_MEGARA_orbitas}]{Carrera2021}.
Thanks to the work of \citet{Antoja2021} based on \textit{Gaia} EDR3 data \citep{Gaia2021}, we know that the orbits of other two outer disc OCs, Saurer~1 and Berkeley~29, are also typical of the thin disc objects. 

\begin{figure}
    \centering
    \includegraphics[width=\columnwidth]{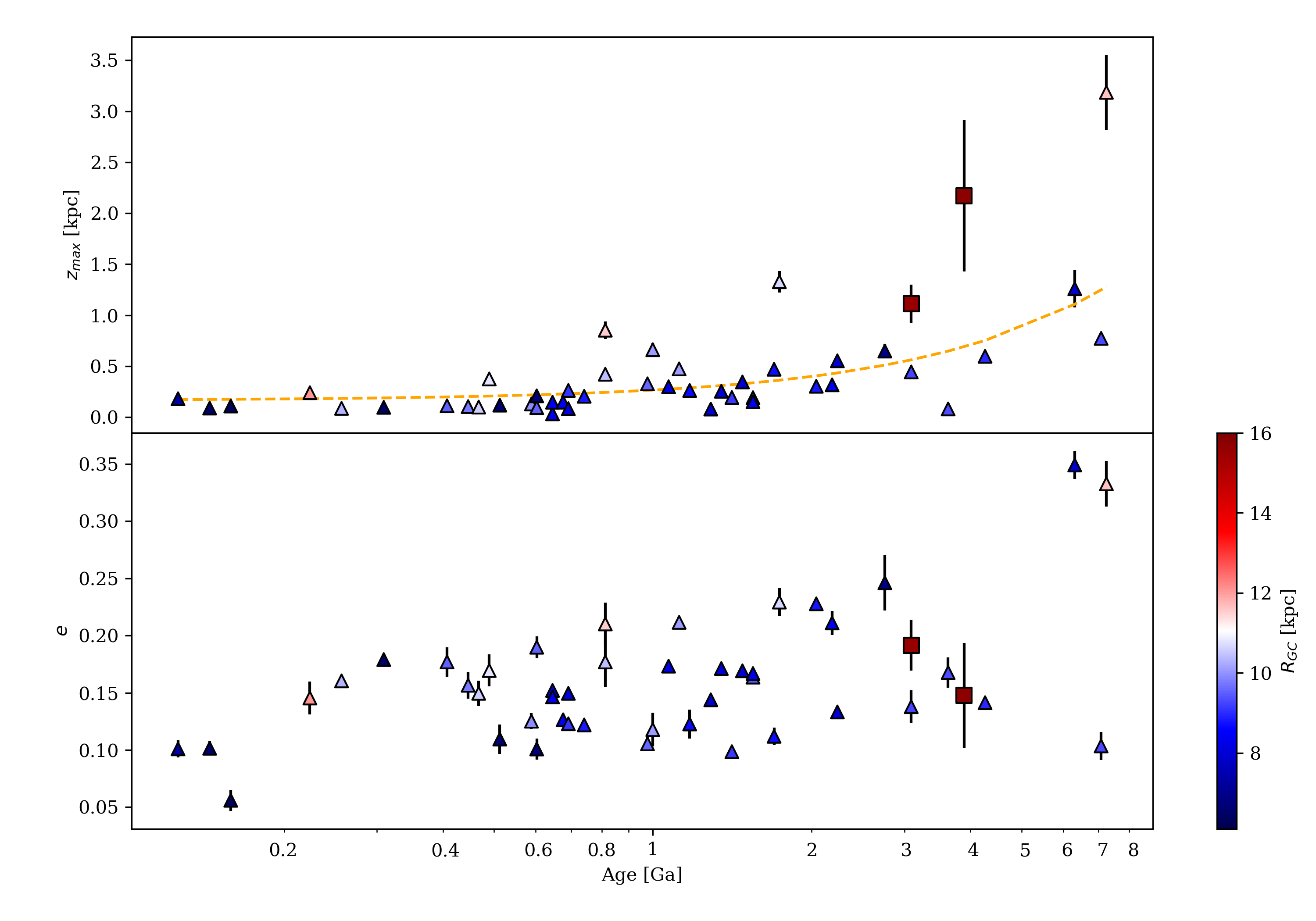}
    \caption{Dependence of the orbital parameters $z_{\rm max}$ (top) and eccentricity (bottom) on the age, and Galactocentric radius of OCs. In the top panel, we fit the data with an exponential function. Objects observed with MEGARA are marked with circles, and those with OCCASO \citep{Carrera2021} with triangles. In all of them, the Galactocentric radius is colour-coded.}
    \label{fig:zmax_e-age_MEGARA}
\end{figure}

Figure~\ref{fig:zmax_e-age_MEGARA} shows the dependence of $z_{\rm max}$ (top) and eccentricity (bottom) on age and $R_{GC}$ (colour-coded). The orbits have been calculated with the potential MW2014 + bar + arms. In this figure, we compare the results of Auner~1 and Berkeley~102 with those of the OCCASO sample \citep{Carrera2021}.  
Similar results are obtained using only the MW2014 potential and including the bar and spiral arms separately. 
Regardless of the chosen potential, the two clusters observed with MEGARA have higher values of $z_{\rm max}$ than OCs of similar age. This is because they are in the outer disc, where the Galactic potential is weaker and, therefore, the orbits reach higher $z_{\rm max}$ values more easily \citep{cantat-gaudin2020}. 
The eccentricity of the studied clusters in the outer disc is within the range of eccentricities expected for their age when compared to the OCCASO sample.  


\section{Abundance determination}\label{Sec:MEGARA_AP-ABU}

We use spectral synthesis with \texttt{iSpec} for the determination of atmospheric parameters and chemical abundances. The chosen radiative transfer code is SPECTRUM, and we use MARCS stellar atmospheric model. We used the sixth version of the GES line list \citep{heiter2021}, and determine our own solar abundances, as shown in next sections.     

    \subsection{Line list selection and atmospheric parameters}

We select which lines are usable at the resolution and spectral range of the instrument to derive both the atmospheric parameters and chemical abundances. We discard lines with a shallow depth compared to the noise (less than 2$\sigma$) in the normalised spectrum of the star Auner\,1\_3, that has the highest S/N.
This initial list has been refined, selecting unblended lines at medium resolution by comparing synthetic spectra with the atmospheric parameters of the observed stars with and without the line of interest, as done in \citet{heiter2021}.
In addition, we checked the abundances in the six analysed stars, discarding lines with systematically discrepant abundances. In Table~\ref{tab:line_list_MEGARA} we show the final selection of lines. 
We have not used the flags of \citet{heiter2021} as a selection criterion due to the small number of lines available in the spectral region studied.


In order to derive the atmospheric parameters, we should take into account that MEGARA in its HR-R highest-resolution mode, only a small spectral window is accessible, making it impossible to achieve a high precision in the measurement of effective temperature and surface gravity. Therefore, we have obtained these parameters from the \texttt{StarHorse} catalogue \citep{Anders2022}, as mentioned before (see Table \ref{tab:MEGARA_stars}). 
In addition, the line broadening is produced by macroturbulence velocity, stellar rotation and the resolving power of the instrument. Both stellar broadening mechanisms are difficult to discern in red giants. \citep[e.g.][]{thygesen2012}. We decided to use only one free parameter, stellar rotation, to perform the fit. Due to this approximation, the obtained value is not physical, so, we do not publish it.
The technique used in the analysis is spectral synthesis by means of \texttt{iSpec} \citep{blancocuaresma2019}.
By fitting all the lines in our spectral window (see Table \ref{tab:line_list_MEGARA}), we determine the metallicity, and microturbulence. 
The process is performed by comparing the observed spectrum with synthetic spectra, and then a least-squares algorithm minimizes the differences (i.e., by computing $\chi^2$) by varying the atmospheric parameters until convergence is reached.
The measured spectroscopic atmospheric parameters are shown in Table~\ref{tab:MEGARA_abundances}. 

\subsection{Chemical abundances:}
Once the atmospheric parameters are determined, the abundances $A(X)$ of individual elements are calculated using spectral synthesis with \texttt{iSpec}. For each chemical species, the abundance is computed as the mean of the studied lines, weighted by the uncertainties. The associated uncertainty of the mean is determined as the standard deviation divided by the square root of the number of lines. 
For elements where only a single line is measured (Co, Ba, and Eu), the uncertainty of that line is adopted as the elemental abundance error. 
In our experience, when iSpec uses a single line to determine abundances, it overestimates the error, which is particularly evident in the Eu measurement. When looking at the appendix figures \ref{fig:Be1021} \ref{fig:Be1022} \ref{fig:Be1023} we clearly see an absorption line of Eu in two stars of Berkeley\_102. 
However, in the other stars, Eu is at the detection limit for the S/N of the spectra obtained.
The chemical abundances are presented in Table~\ref{tab:MEGARA_abundances}. We compute the abundances of Ca, Co, Ni, Ba and Eu, not being able to determine the abundance of the last element in any star of Auner~1.  
Abundance values relative to the Sun ([X/H]) were calculated, with solar abundances determined from sky spectra taken during twilight. Table~\ref{tab:solar_MEGARA} provides the solar values obtained in comparison with the literature. Solar reference values were calculated using the same procedure applied to the stars under analysis. Differences observed for certain elements compared to the literature \citep{Grevesse2007, Asplund2009} are primarily attributed to the use of different lines between studies. 

The average uncertainties obtained for the abundance of each star are: 0.04\,dex for Fe, 0.11\,dex for Ca and Ni, 0.61\,dex for Co, 0.37\,dex for Ba and 1.55\,dex for Eu. It is important to notice that  when there is only one line, \texttt{iSpec} tends to give very large errors, as  discussed in \citet{Casamiquela2020}. Therefore, the uncertainties of the elements Co, Ba, and Eu may be overestimated.  
Nevertheless, in order to study the Eu accurately, it would be necessary to obtain observations with higher S/N, and therefore longer exposure times.     

\subsection{Abundance zero-point corrections}
In order to quantify the abundance biases at play in our study, we have analysed eleven stars in common between APOGEE and MEGASTAR DR1.1\footnote{https://www.fractal-es.com/
megaragtc-stellarlibrary} project \citep{Molla2023}, observed with the same instrumental configuration of MEGARA, the HR-R VPH and the IFU mode. We selected the stars with \teff  \,and \logg \,of RC and RGB stars and analysed them with the same procedure as described for cluster stars. Figure~\ref{CaFe} shows the comparison between [Fe/H] obtained from MEGASTAR and APOGEE spectra in the APOGEE-MEGASTAR direction.
This allowed us to identify a systematic difference of our results of $\Delta$[Fe/H]=$-0.21\pm$0.05\,dex from those of APOGEE. We have not found an obvious reason for this bias, and we guess that it may be due to the reduced spectral range and resolution of the MEGARA in comparison with those of APOGEE, and the way in which the atmospheric parameters were calculated.

We have also studied the differences $\Delta$[X/Fe] for the other elements in common with APOGEE, and their comparisons are shown also in the Fig. \ref{CaFe} of the appendix. Ba and Eu are not included in this comparison because APOGEE has not studied them. The resulting means of the differences are as follows: $\Delta$[Ca/Fe]=$-0.05\pm$0.07\,dex, $\Delta$[Co/Fe]=0.19$\pm$0.14\,dex,
$\Delta$[Ni/Fe]=$-0.02\pm$0.03\,dex.
In all these cases the differences are similar or smaller than the uncertainty.
The mean differences found for all the elements are applied as a correction offset.
The final abundances used in this work are shown in Table~\ref{tab:MEGARA_abundances}.

\subsection{Final cluster abundances}
The abundance of each cluster is calculated as the weighted average of the abundances of its member stars. The associated uncertainty is the standard deviation of the abundances taking into account the low statistics correction factor in $\sigma$ by applying the equation~5 of \citet{Roesslein2007}, as done in the computation of $v_{\rm scatter}$.      
The abundances of each cluster are listed in Table \ref{tab:MEGARA_ocs}. The average standard deviations we found for the clusters are: 0.09\,dex for Fe, 0.17\,dex for Ca, 0.18 dex for Ti,  0.14\,dex for Co, 0.07\,dex for Ni, 0.12\,dex for Ba, and 0.05\,dex for Eu. 

There is only one measure of [Fe/H] abundance for Berkeley~102 in the literature, that provided by \citet{Perren2022}. The authors determined the [Fe/H] by fitting {\it Gaia} EDR3 photometry to synthetic cluster with a Bayesian inference algorithm. They obtained a value of $-0.17^{0.07}_{-0.45}$\,dex compatible with our value taking into account the uncertainties. Since we use medium-resolution spectroscopy, our result has a higher precision than the one obtained in the cited work. 
The metallicity of Auner~1 had not been measured photometrically nor spectroscopically before.

\begin{table*}[h!]
\setlength{\tabcolsep}{0.6mm}
\begin{center}
\caption{Spectroscopic properties of the studied stars.}
  \begin{tabular}{lcccccccccc}
  \hline
Name &    [M/H] &    $v_{\mathrm{mic}}$ &    [Fe/H] &[Ca/Fe] &[Ti/Fe] &[Co/Fe] &[Ni/Fe] &[Ba/Fe] &[Eu/Fe]\\
 & [dex] &[\kms] &[dex] &[dex] &[dex] &[dex] &[dex] &[dex] &[dex]\\
\hline
Auner\,1\_1 &    -0.43 $\pm$ 0.31 &   1.3 $\pm$ 0.57 &  -0.12 $\pm$ 0.08 & -0.20 $\pm$ 0.35 & -- & --  & 0.08 $\pm$ 0.22 &   0.24 $\pm$ 0.93 &     -- \\
Auner\,1\_2 &    -0.38 $\pm$ 0.26 &   1.3 $\pm$ 0.49 &  -0.33 $\pm$ 0.03 &  -0.05 $\pm$ 0.07 & -0.06$\pm$0.12 &  0.24 $\pm$ 0.72 &   0.13 $\pm$ 0.11&  0.31 $\pm$ 0.92 &     -- \\
Auner\,1\_3  &   -0.25 $\pm$ 0.14 &  1.29 $\pm$ 0.22 &  -0.28 $\pm$ 0.03 &   0.20 $\pm$ 0.02 & 0.11$\pm$0.20 & -0.10 $\pm$ 0.72  &  -0.05 $\pm$ 0.13 &  0.09 $\pm$ 0.97 &     --\\
Be\,102\_1  &  -0.74 $\pm$ 0.14 &  1.66 $\pm$ 0.28 & -0.42 $\pm$ 0.03 &-0.08 $\pm$ 0.08 & -0.07$\pm$0.07 &  -0.05 $\pm$ 1.03  &  -0.06 $\pm$ 0.09 &  0.15 $\pm$ 1.07 &   0.67 $\pm$ 1.30\\
Be\,102\_2  &  -0.60 $\pm$ 0.18 &  1.77 $\pm$ 0.29 & -0.32 $\pm$ 0.04 & 0.22 $\pm$ 0.02 & -0.17$\pm$0.06 &  0.07 $\pm$ 0.78  &  0.05 $\pm$ 0.06 &   0.31 $\pm$ 0.93 &   0.58 $\pm$ 1.79\\
Be\,102\_3  &  -0.34 $\pm$ 0.19 &  1.48 $\pm$ 0.32 & -0.26 $\pm$ 0.04 & 0.12 $\pm$ 0.07 & 0.01$\pm$0.20 & -0.14 $\pm$ 1.01  &   -0.04 $\pm$ 0.17 &  0.44 $\pm$ 0.76 &  -- \\   
     \hline
  \end{tabular}
\label{tab:MEGARA_abundances}
\end{center}
\end{table*}

\section{Galactic abundance trends}\label{OPLUS}

In this section, we discuss the chemical abundances of Auner~1 and Berkeley~102 in the broader context of Galactic chemical evolution.
For this purpose, we compare our results with those of the OCCASO+ sample \citep{Carbajo2024}, a compilation of 99 clusters that have spectroscopic observations of at least 4 red giant stars at resolution $R$ > 20\,000, made from observations of the OCCASO \citep{Carbajo2024}, GES DR5 \citep{Magrini2022}, APOGEE DR17 \citep{myers2022} and GALAH DR3 \citep{spina2021} surveys. See Fig. \ref{fig:Fe_Rgc}.

\subsection{Abundance trends with [Fe/H]}
    \label{sect:metallicity}

\begin{figure*}
\centering
\includegraphics[width=2\columnwidth]{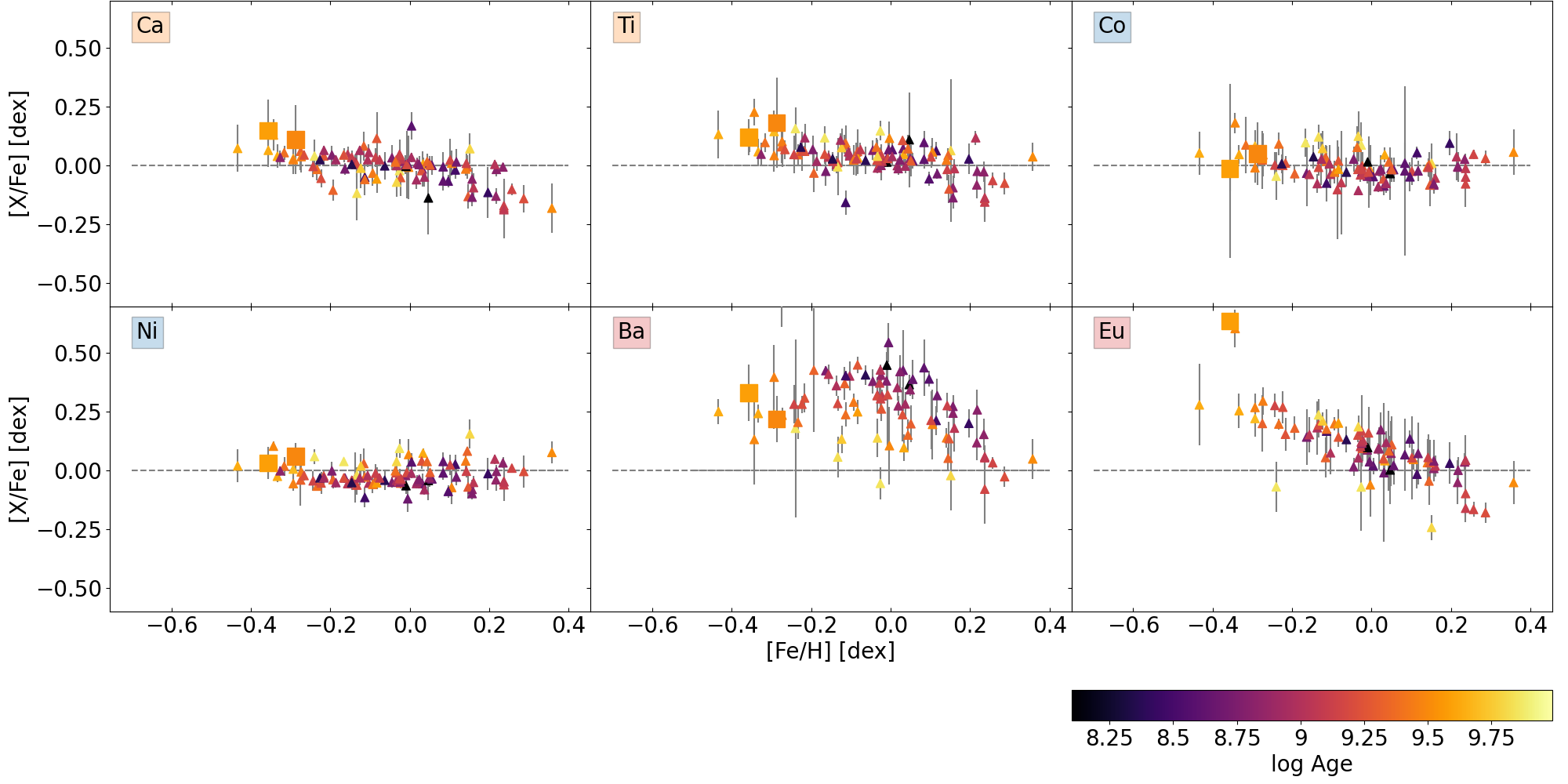}

\caption{Dependence of [X/Fe] on [Fe/H]. OCCASO+ data \citep{Carbajo2024} are represented by triangles. Auner~1 and Berkeley~102 are marked as squares. All objects are colour-coded by age.}
\label{fig:FeH_M}
\end{figure*}

In Fig.\ref{fig:FeH_M} we show the results of [X/Fe] on [Fe/H] of the two clusters in relation with the OCCASO+ sample. There is a slightly decreasing trends of the $\alpha$ elements [Ca/Fe] and [Ti/Fe] with [Fe/H], as widely reported in the literature. These trends are explained by the production of $\alpha$ elements mainly in core collapse supernovae (CCSs) from massive stars in short timescales in comparison with Fe, which is produced on longer timescales mostly by type Ia Supernovae (SNe Ia).
The abundances of [Ca/Fe] and [Ti/Fe] of both clusters are slightly higher than those of OCCASO+ for this [Fe/H] abundance, but compatible taking into account the uncertainties.

The Fe-peak elements (Ni and Co) are thought to be produced by the same processes as Fe \citep{kobayashi2020}, with a trend that is generally flat. The values of [Ni/Fe] of both clusters are compatible with those of OCCASO+. In \citet{Carbajo2024} a  mild decreasing trend was described at low metallicity for Co. Even though, the values of Auner~1 and Berkeley~102 do not follow such trend, they have compatible values, taking into account uncertainties, though. 

Neutron capture elements can be produced by slow (s) or fast (r) processes of neutron capture. They are defined by whether the capture timescale is longer or shorter than $\beta$ decay, and occur at different astrophysical sites. 
Ba is mostly produced by s-process in AGB stars \citep[e.g.][]{Gallino1998}, and shows a large dispersion due to a larger uncertainty and a dependence on age, being younger OCs more enhanced. [Ba/Fe] shows a slight increasing trend with [Fe/H], reaching their maximum at [Fe/H] $\sim$ 0\,dex to decrease again more abruptly at higher [Fe/H] abundances. This last decrease is explained since as [Fe/H] increases, the ratio of neutrons to Fe in AGB star decreases. As a result, there is a smaller proportion of s-process elements being produced \citep{Gallino2006,Cristallo2009,Karakas2014}. The abundances of both clusters have [Ba/Fe] abundances compatible with those of the OCCASO+ sample. 

The origin of the r-process is still under debate \citep[e.g.][]{Kajino2019}, but for it to occur, the neutron density and temperature must be high. Several scenarios have been proposed for its production in CCS, such as neutrino-induced winds \citep{Woosley1994}, or rotating CCS polar jets \citep{Nishimura2006}. Neutron star collisions \citep{Freiburghaus1999} and neutron star-black hole mergers \citep{Surman2008} have also been proposed, among others. The collision of neutron stars is a scenario that has been favoured lately due to the identification of Sr lines in the kilonova AT2017gfo \citep{Watson2019}, a product of the collision of two neutron stars. The timescale of production per r-process will depend on the physical object in which it is produced and it is unknown. [Eu/Fe] shows a steeper decreasing slope with [Fe/H] compared to [Ca/Fe], which denotes a rapid production of this element compared to Fe. Berkeley~102 shows a higher [Eu/Fe] value, albeit with a large intrinsic uncertainty in the determination of abundances of this element, $\sim$1.5\,dex, that renders this measurement still 1$\sigma$-compatible with the general trend. It is interesting to notice that the [Eu/Fe] abundance is similar to that of Berkeley~29, the outermost cluster in the global sample.


\subsection{Radial trends}\label{sect:radial_trends}

\begin{figure}
\centering
\includegraphics[width=\columnwidth]{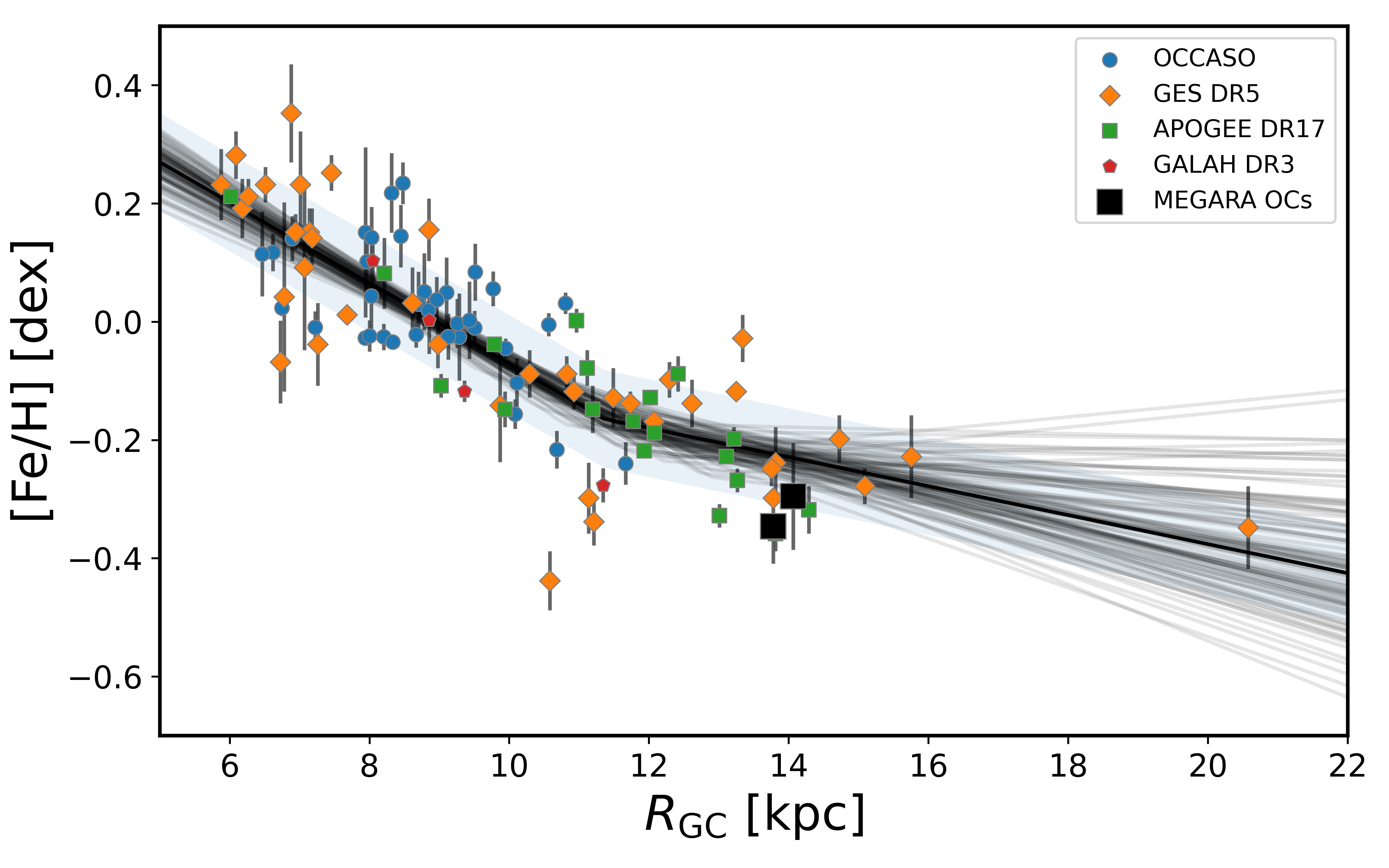}
\caption{Dependence of [Fe/H] on the Galactocentric radius. The different sub-samples of OCCASO+ as in Fig.~14 \citet{Carbajo2024} are colour-coded. Auner~1 and Berkeley~102 are marked as black squares. The grey lines are all the fits made using MCMC and the black line shows the best fit.}
\label{fig:Fe_Rgc}
\end{figure} 

In Fig.~\ref{fig:Fe_Rgc}, we place Auner~1 and Berkeley~102 in the [Fe/H] vs. Galactocentric radius diagram, together with the clusters of OCCASO+. 
We perform a linear fit to quantify the radial gradient. This fit was performed by the same method as in \citet{Anders2017}, using a maximum likelihood algorithm as first guess, and computing a Markov-Chain Monte-Carlo (MCMC) with the python package \emph{emcee} \citep{Foreman2013}.
The grey lines in the figure are all the fits made with MCMC and the black line shows the best fit.
The computed slope inside the knee radius is -0.068$\pm$0.007\,dex\,kpc$^{-1}$, and outside the knee radius is -0.025$\pm$0.011\,dex\,kpc$^{-1}$. We find the knee position at 11.4$\pm$0.8\,kpc. The fit is in agreement with that obtained in \citet{Carbajo2024, myers2022, Magrini2022}.
Both Auner~1 and Berkeley~102 have metallicities slightly below the general trend, but they are compatible with other clusters at similar Galactocentric radii. 
The four outermost clusters (Berkeley~21, Berkeley~31, Tombaugh~2 and Berkeley~29) have much relevance to this fit, since fitting without them, we continue to find the knee, although the second slope is compatible with the first one taking into account the uncertainty. The abundances we observe for Auner~1 and Berkeley~102, denote that the knee position may be further away from what we find when evaluating the clusters studied in the sample that we present, and that the flattening may only occur in clusters beyond 14\,kpc. 

\begin{figure*}
\centering
\includegraphics[width=2\columnwidth]{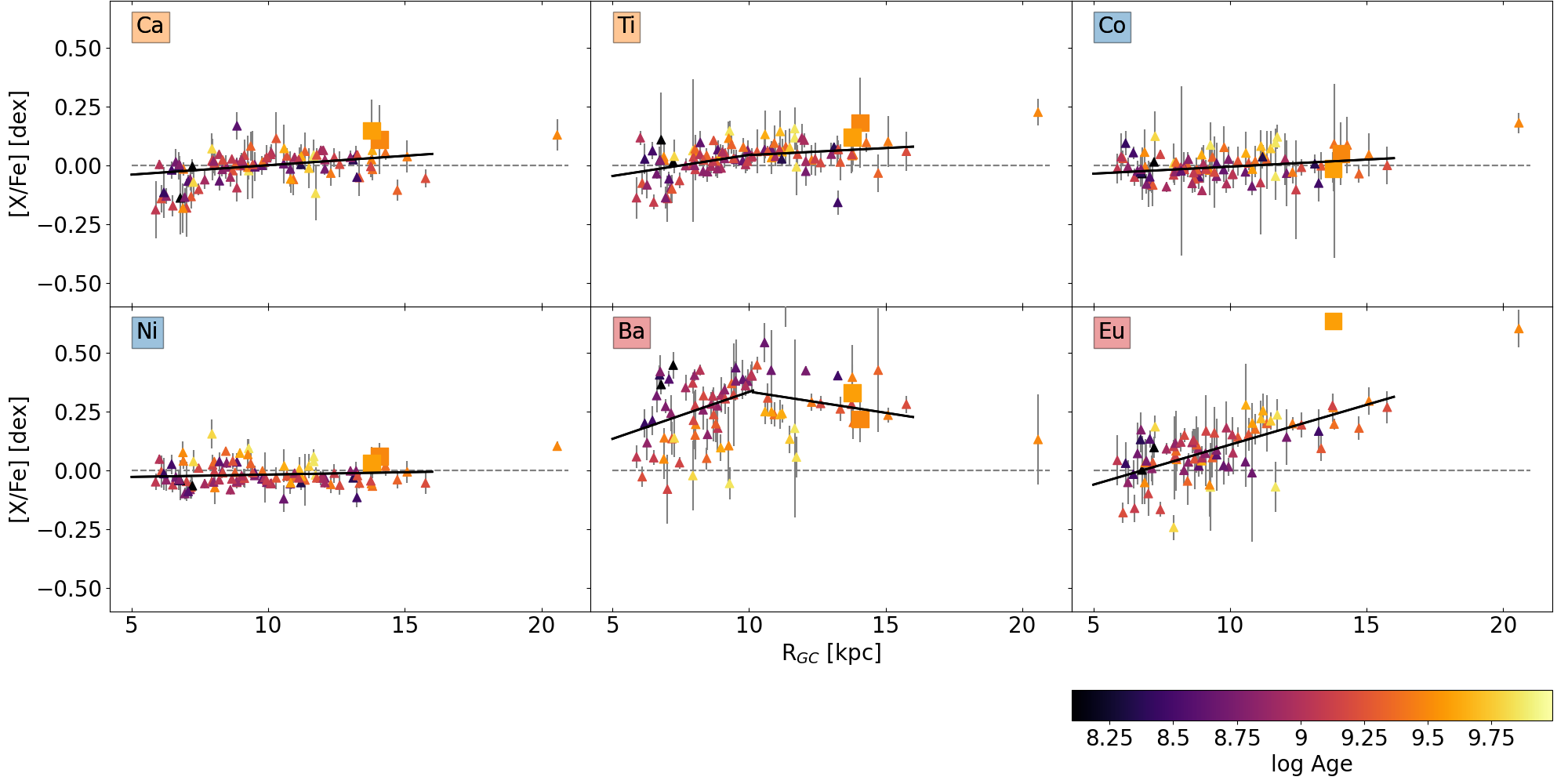}

\caption{Dependence of [X/Fe] on Galactocentric radius. OCCASO+ data \citep{Carbajo2024} are represented by triangles. Auner~1 and  Berkeley~102 are marked as squares. All the objects are colour-coded by age. The black lines are the MCMC best fits to the OCCASO+ sample as derived by \citet{Carbajo2024}.}
\label{fig:Rgc_M}
\end{figure*}

When examining the radial trends of other elements with respect to iron (Fig.~\ref{fig:Rgc_M}), the trends are analysed using MCMC as outlined in the previous paragraph. The algorithm employed determines whether a single slope or a combination of two slopes provides a better fit. The analysis shows that all elements are best described by a single slope, except for Ba, which is better described by two.
The $\alpha$ elements [Ca/Fe] and [Ti/Fe] show a mild positive gradient, in agreement with the models of the inside-out formation of the Galactic disc.
The Fe-peak elements [Co/Fe] and [Ni/Fe] do not present radial gradients due to their production by the same processes as Fe.  
Inside the knee ($R_{\rm Gal}\leq11.4$\,kpc), [Ba/Fe] shows a positive trend, which is reversed in the outer disc. This pattern is produced by the dependence of s-process production on [Fe/H], as previously discussed.
The [Eu/Fe] has a steep slope, which is consistent with its production mainly by r-process.
We also see that in general, both clusters have abundances compatible with those of the other clusters with similar Galactocentric radii. In concordance with Fig.~\ref{fig:FeH_M}, the [Ca/Fe], [Ti/Fe] and [Eu/Fe] values for our clusters are slightly higher than the general trend, but still compatible within the $1\sigma$ uncertainties. As mentioned above, the [Eu/Fe] abundance in Berkeley~102 cluster is similar to that of Berkeley~29, located at a larger Galactocentric radius. But again, we remark the large intrinsic uncertainties in our measurements for this element.
In the other elements, the abundances of both clusters coincide perfectly with those found in the OCCASO+ sample \citep{Carbajo2024}.   

\subsection{Azimuthal dependence}\label{sect:Azimuthal_dependence}

\begin{figure*}
\centering
\includegraphics[width=0.9\textwidth]{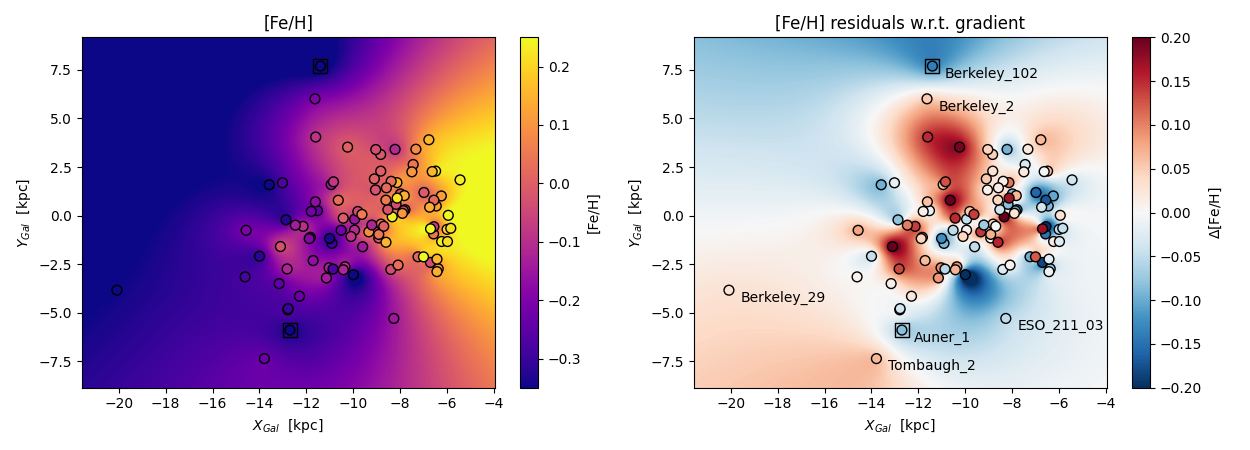}
\caption{Kriging maps of the metallicities of OCs in the Galactic plane. Left: [Fe/H] abundances. Right:
Residuals once the radial gradient (Fig. \ref{fig:Fe_Rgc}) is subtracted. In both panels, we show the OCCASO+ sample \citep{Carbajo2024}. The Sun is situated at ($-8.2, 0.0$) in both maps. The MEGARA clusters Auner~1 and Berkeley~102 are marked as squares, and some other distant clusters included in OCCASO+ are also labelled.}
\label{fig:Az_Trend}
\end{figure*}

Figure~\ref{fig:Az_Trend} shows the Kriging maps\footnote{\url{http://pykrige.readthedocs.io/}} of the metallicities of OCs observed by high-resolution spectroscopy in the Galactic plane. In particular, the right panel shows the Kriging map of the residuals once the radial gradient is subtracted. In both panels the clusters of the OCCASO+ sample are shown \citep{Carbajo2024}, and the two clusters in this paper are marked with squares. When studying the map of the [Fe/H] residuals shown in the right panel of Fig.~\ref{fig:Az_Trend}, we find that there is a certain structure visible that is not produced by the metallicity gradient. The pattern is similar to that found in \citet{Poggio2022} in the regions in which our study and theirs overlap (i.e., within 4\,kpc from the Sun). The authors explained this variation as produced by the formation of the stars in the region of the spiral arms or in the inter-arm region. We agree with \citet{Poggio2022} in the explanation. This dependence is partially responsible for the dispersion of [Fe/H] for the same $R_{\rm GC}$.    
These maps feature well-sampled regions, primarily within a 3\,kpc radius of the Sun. However, it is important to note that some areas are dominated by a single cluster. As such, these maps are best suited for guiding our interpretation and planning future observing campaigns. Clearly, more observations of distant, both in Galactocentric distance and azimuth, open clusters with medium-to-high resolution instruments are needed to firmly establish the abundance trends of the outermost parts of the Galactic disc. Since our statistics are small, it is important to compare the relatively sparse open cluster sample with other abundance tracers that are visible at such distances, such as Cepheids or red-giant stars. 

\subsection{Implications for the formation of outer-disc OCs}

Galactic chemical evolution models based on the inside-out formation scenario often predict a continuously decreasing radial gradient \citep[e.g.][]{prantzos2018,kobayashi2020}. The observed flattening can be matched when including extra pre-enriched infall of material in the outer disc in Galactic chemical evolution models \citep{magrini2009}. Similar patterns in metallicity are found in extragalactic studies based on ionized gas abundances \citep{bresolin2009, bresolin2012} . Resonances of bar or spiral arms \citep{lepine2011} or accretion of gas from the intergalactic medium have been proposed as mechanisms to explain this feature \citep{bresolin2012, bowen2020}. In a third scenario, the outer disc OCs are associated with merger events, either acquired from merging galaxies \cite{Friel2013} or formed as a result of the merger event itself \citep{Yong2012, magrini2009}. Since the enrichment process in this zone would be different, a peculiar signature of abundances beyond metallicity is expected, but not observed \citep{Carraro2007, Sestito2008}. The orbits of the OCs could also be affected by the trajectory of the merger galaxy and the reaction of the disc. Another possible scenario is that these clusters have formed in a more internal region of the disc and have migrated outwards, producing the flat abundance gradients. This process has been simulated  \citep{Roskar2008} and proposed to explain the observed age-[Fe/H] distribution of outer disc field stars \citep{lian2022}. 
In this case, the chemical abundances of the outer disc OCs should be the ones of innermost regions at the time of its birth.
Based on our analysis of age, orbital parameters, and chemical abundances, we conclude that Auner~1 and Berkeley~102 are part of the thin disc population. This evidence strongly challenges the hypothesis that the outermost open clusters originated in another galaxy and were later accreted during a merger event. The flattening pattern found in Fe in the outermost part of the disc is dominated by four open clusters observed by GES. Further research is needed in this region.

\section{Summary and conclusions}
\label{sec:conclusions}

In this study, we have used medium-resolution spectra ($R \sim 18\,700$) obtained with the MEGARA instrument at the GTC to study red-giant star members of the old and distant outer-disc clusters Auner~1 and Berkeley~102. This is the first time that these clusters have been studied spectroscopically, and, to our knowledge, the first time that open clusters at such Galactic azimuths and radius have been observed. There are only a few clusters spectroscopically studied at the distance at which these OCs are located, being beyond the capabilities of most telescopes/instruments due to the faintness of the studied stars ($G\geq16$). We determine
radial velocities and atmospheric parameters for the member stars, as well as updated ages and distances for the two clusters. Finally, we have measured the chemical abundances of six chemical elements (Fe, Ca, Co, Ni, Ba, and Eu) using spectral synthesis, differentially correcting for abundance offsets with respect to high-resolution spectroscopy.

Our main findings can be summarised as follows: 
\begin{enumerate}
    \item The radial velocities of the clusters and their orbits confirm that both OCs have kinematics typical of the outer Galactic disc.   
    \item The [Fe/H] values of Auner~1 and Berkeley~102 are following the trend of other outer-disc clusters located at  similar Galactocentric radii, with hints for systematic azimuthal abundance variations, as also found in field-star studies.
    \item The [Fe/H] abundances of  Auner~1 and Berkeley~102 suggest that the knee position of the radial gradient may be sightly further away than recent estimates (e.g. 11.2\,kpc; \citealt{Carbajo2024}).
    \item The [X/Fe] abundances follow the expected trends for the Galactocentric distances of Auner~1 and Berkeley~102, and are compatible with the OCCASO+ sample. Slightly higher values are found for [Ca/Fe] and [Eu/Fe], although still compatible within $1\sigma$. 

\end{enumerate}

From the results of our age, orbit, and chemical-abundance analysis, we conclude that Auner~1 and Berkeley~102 belong to the thin disc population. 
Thanks to the work of \citet{Antoja2021} based on \textit{Gaia} EDR3 data \citep{Gaia2021}, we know that the two outermost known OCs of the Milky Way, Saurer~1 and Berkeley~29, are also very likely thin disc objects. These facts combined provide strong evidence against the scenario of the outermost OCs having formed in another galaxy and being captured during a merger event.  
The data obtained so far do not allow us to differentiate between different outer-disc chemical-evolution and migration scenarios proposed in the literature. 
In order to better characterize the knee position and differentiate between these scenarios, it is essential to study more clusters in the outer disc, and the capabilities of MEGARA at GTC make this possible, opening a window to a largely unexplored area of the Milky Way.


\begin{acknowledgements}
This publication is based on data obtained with the MEGARA instrument at the Gran Telescopio Canarias, funded by European Regional Development Funds (ERDF) through Programa Operativo Canarias FEDER 2014-2020, and installed in the Spanish Observatorio del Roque de los Muchachos of the Instituto de Astrofísica de Canarias, on the island of La Palma. MEGARA has been built by a Consortium led by the Universidad Complutense de Madrid (Spain) and that also includes the Instituto de Astrofísica, Óptica y Electrónica (Mexico), Instituto de Astrofísica de Andalucía (CSIC, Spain) and the Univesidad Politécnica de Madrid (Spain). MEGARA is funded by the Consortium institutions and by GRANTECAN S.A. 
Based on observations made with the Gran Telescopio Canarias (GTC), installed at the Spanish Observatorio del Roque de los Muchachos of the Instituto de Astrofísica de Canarias, on the island of La Palma.
This work is partly based on data from the GTC Public Archive at CAB (INTA-CSIC), developed in the framework of the Spanish Virtual Observatory project supported by the Spanish MINECO through grants AYA 2011-24052 and AYA 2014-55216. The system is maintained by the Data Archive Unit of the CAB (INTA-CSIC)

This work has made use of data from the European Space Agency (ESA) mission \textit{Gaia} (\url{http://www.cosmos.esa.int/gaia}), processed by the Gaia Data Processing and Analysis Consortium (DPAC, \url{http://www.cosmos.esa.int/web/gaia/dpac/consortium}). We acknowledge the \textit{Gaia} Project Scientist Support Team and the Gaia DPAC. Funding for the DPAC has been provided by national institutions, in particular, the institutions participating in the \textit{Gaia} Multilateral Agreement.

This work was supported by the MINECO (Spanish Ministry of Economy, Industry and Competitiveness) through grant ESP2016-80079-C2-1-R (MINECO/FEDER, UE) and by the Spanish MICIN/AEI/10.13039/501100011033 and by "ERDF A way of making Europe" by the “European Union” through grants RTI2018-095076-B-C21 and PID2021-122842OB-C21, and the Institute of Cosmos Sciences University of Barcelona (ICCUB, Unidad de Excelencia ’Mar\'{\i}a de Maeztu’) through grant CEX2019-000918-M. FA and LC acknowledge the grants RYC2021-031683-I and RYC2021-033762-I, funded by MCIN/AEI/10.13039/501100011033 and by the European Union NextGenerationEU/PRTR. 
AGdP acknowledges financial support from the Spanish MICIN under grants PID2022-138621NB-I00 and PID2021-123417OB-I00.

This research has made use of NASA’s Astrophysics Data System Bibliographic Services.

\end{acknowledgements}

\bibliographystyle{aa} 
\bibliography{occasov}

\begin{appendix} 

\section{Complementary tables and figures}

This appendix contains three complementary tables: Table \ref{tab:line_list_MEGARA} shows the linelist used in our analysis of the MEGARA HR-R spectra, Table \ref{tab:Orbitas} lists the orbital parameters of the two clusters obtained under different assumptions, and Table \ref{tab:solar_MEGARA} shows the results of our abundance analysis of a MEGARA spectrum of the Sun.

\begin{figure*}
\centering
\includegraphics[width=0.7\textwidth]{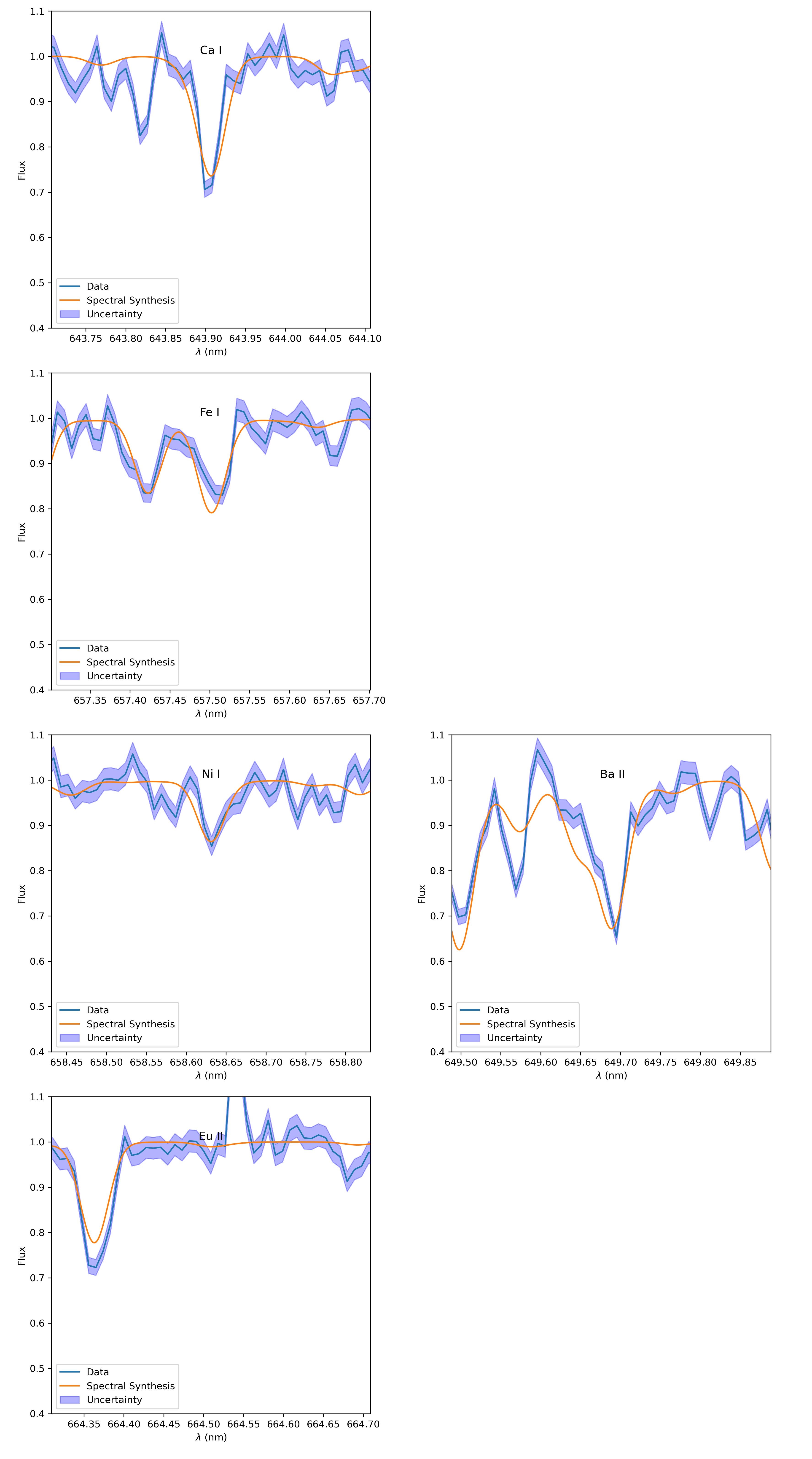}
\caption{Spectrum of Auner\_1\_1 and spectral synthesis fit.}
\label{fig:A11}
\end{figure*}

\begin{figure*}[h]
    \centering
    \includegraphics[width=0.7\textwidth]{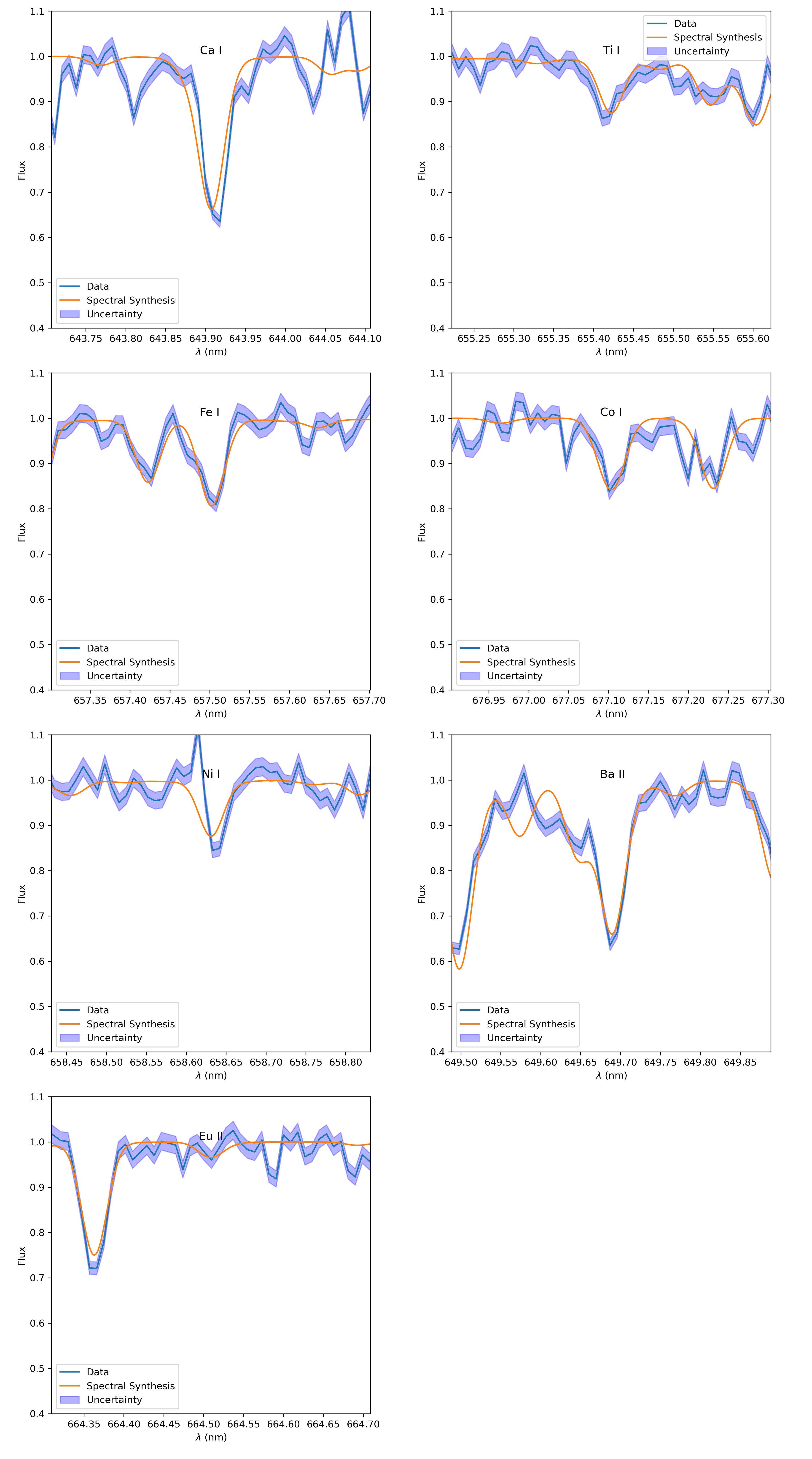}
    \caption{Spectrum of Auner\_1\_2 and spectral synthesis fit.}
    \label{fig:A12}
\end{figure*}

\begin{figure*}[h]
    \centering
    \includegraphics[width=0.7\textwidth]{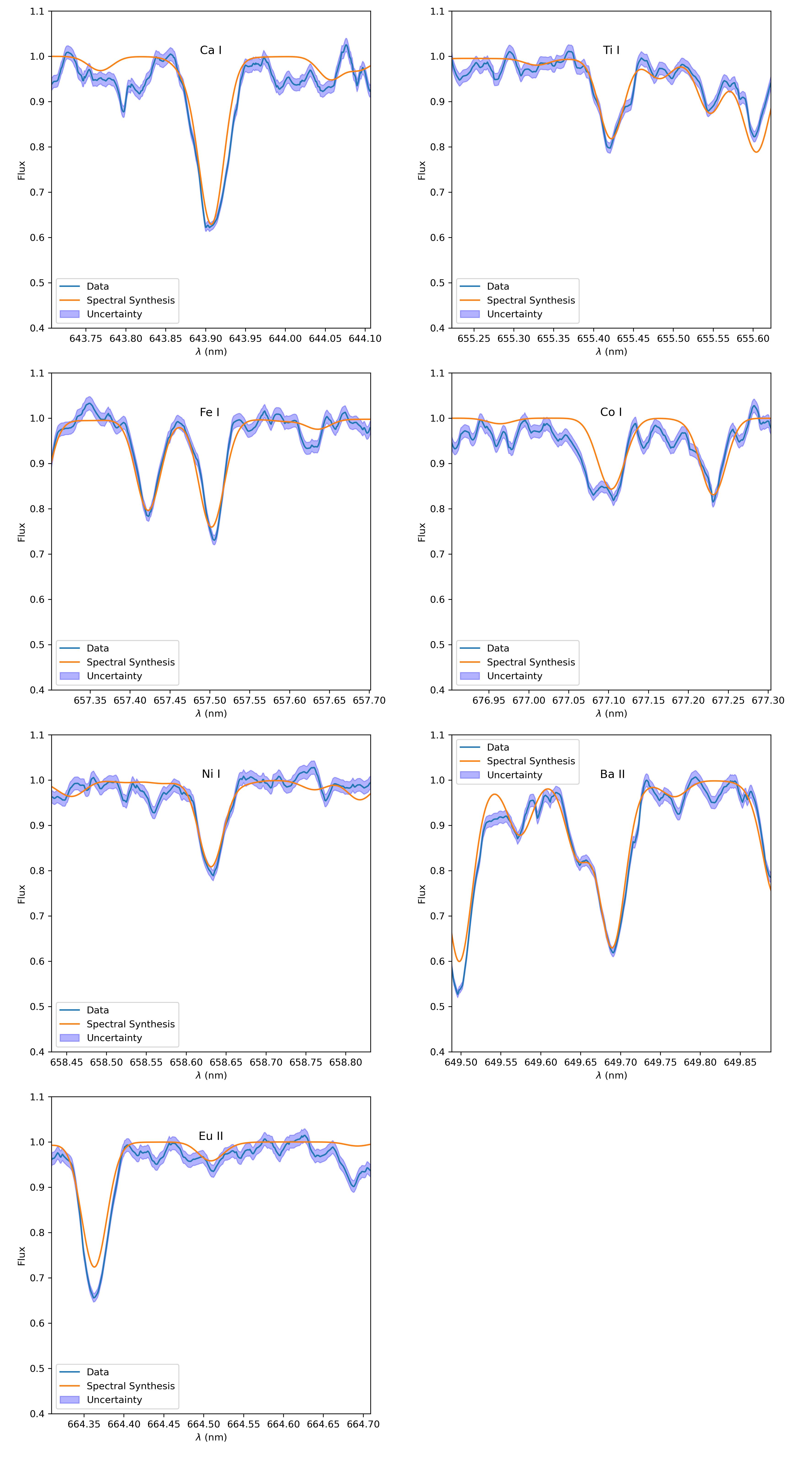}
    \caption{Spectrum of Auner\_1\_3 and spectral synthesis fit.}
    \label{fig:A13}
\end{figure*}

\begin{figure*}[h]
    \centering
    \includegraphics[width=0.7\textwidth]{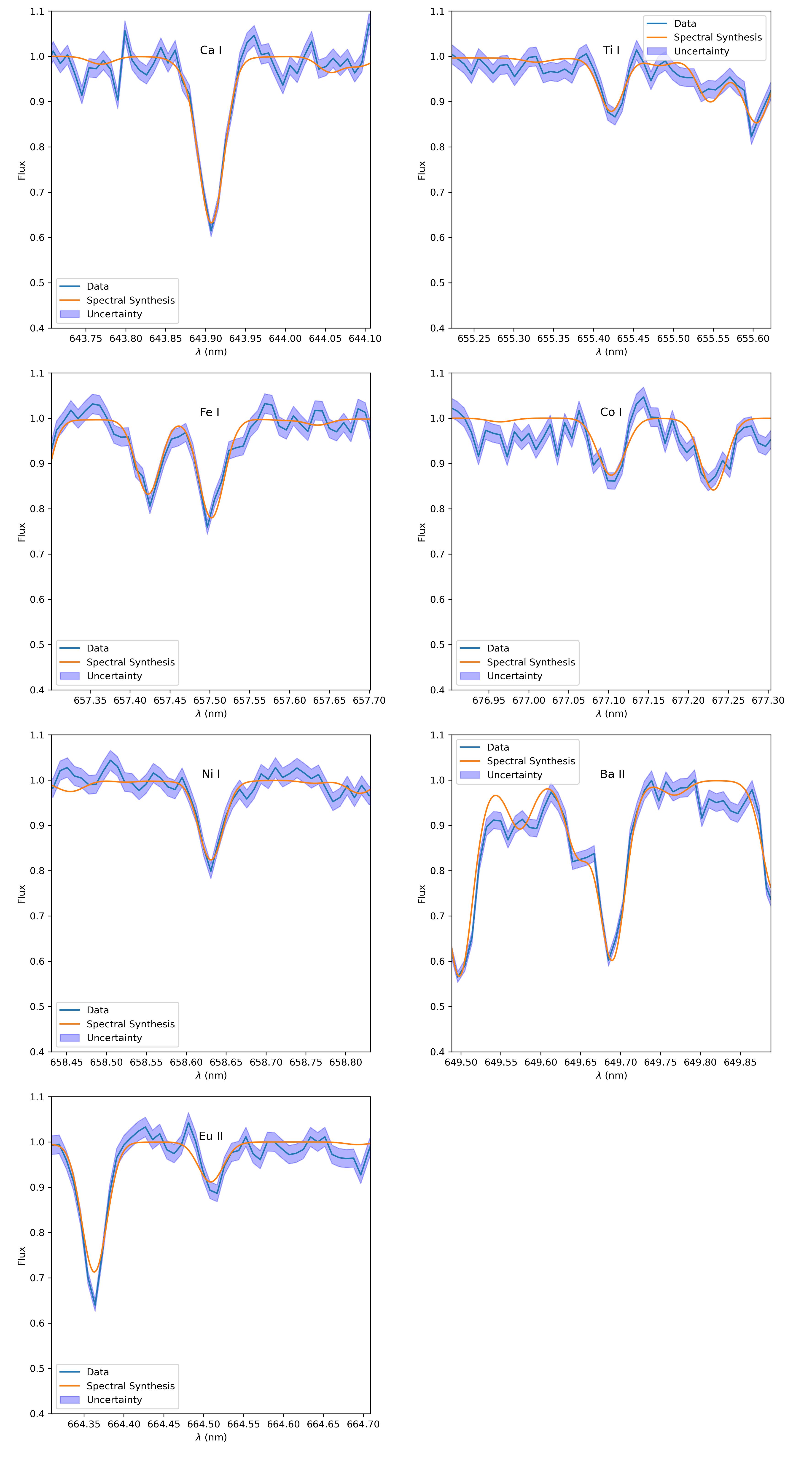}
    \caption{Spectrum of Berkeley\_102\_1 and spectral synthesis fit.}
    \label{fig:Be1021}
\end{figure*}

\begin{figure*}[h]
    \centering
    \includegraphics[width=0.7\textwidth]{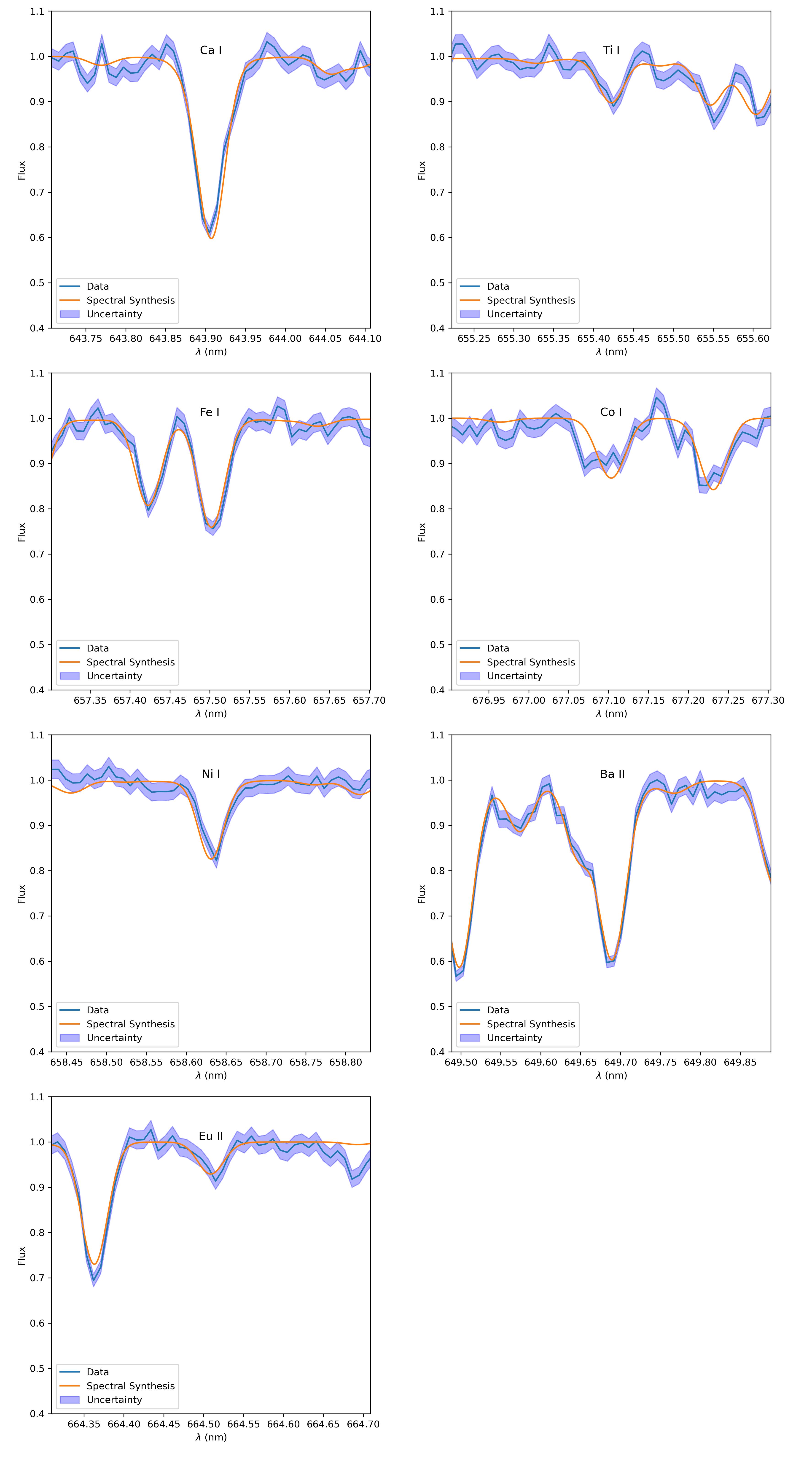}
    \caption{Spectrum of Berkeley\_102\_2 and spectral synthesis fit.}
    \label{fig:Be1022}
\end{figure*}

\begin{figure*}[h]
    \centering
    \includegraphics[width=0.7\textwidth]{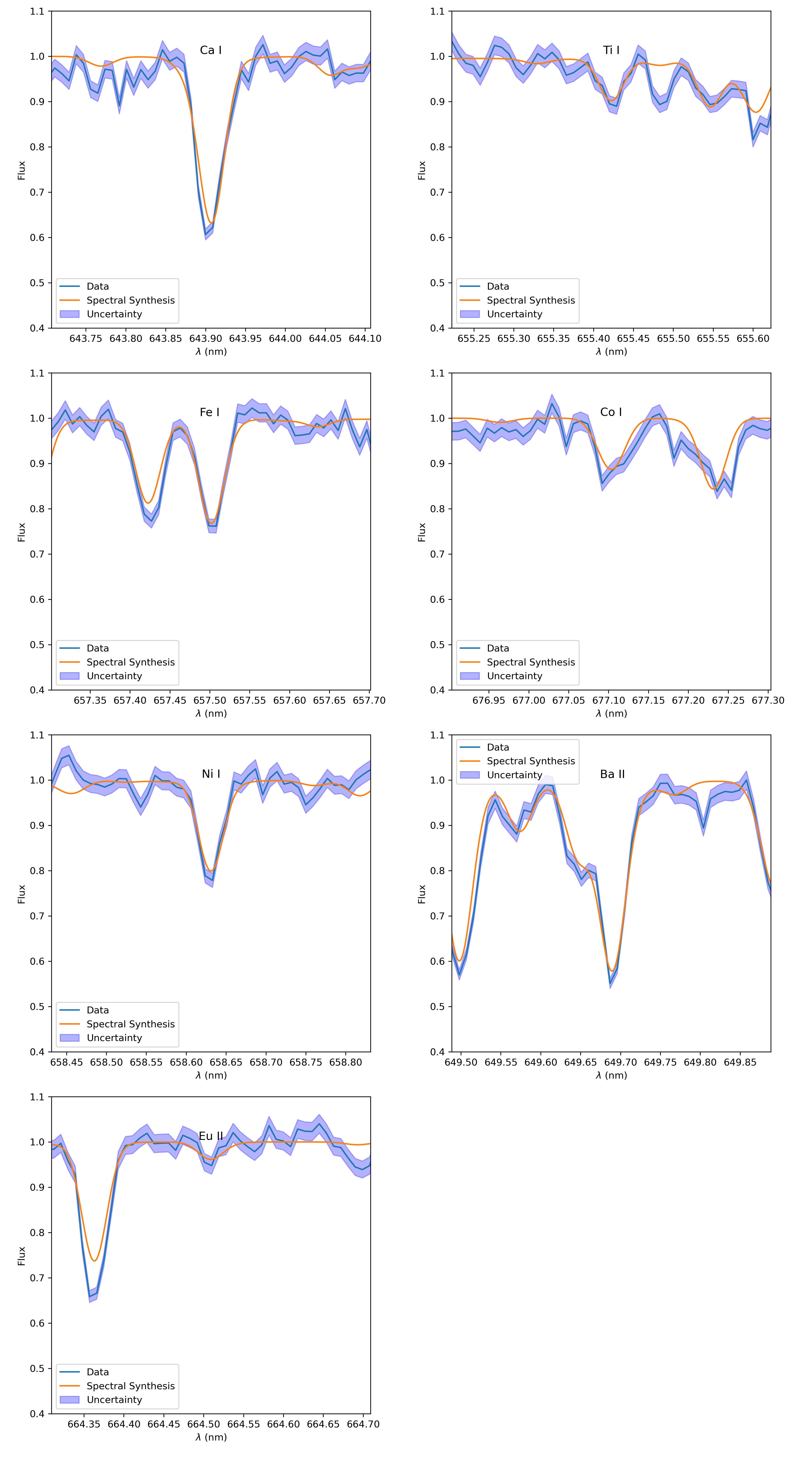}
    \caption{Spectrum of Berkeley\_102\_3 and spectral synthesis fit.}
    \label{fig:Be1023}
\end{figure*}

\begin{figure*}
  \centering
    \includegraphics[width=0.45\textwidth, trim={0 2cm 1cm 3cm}, clip]{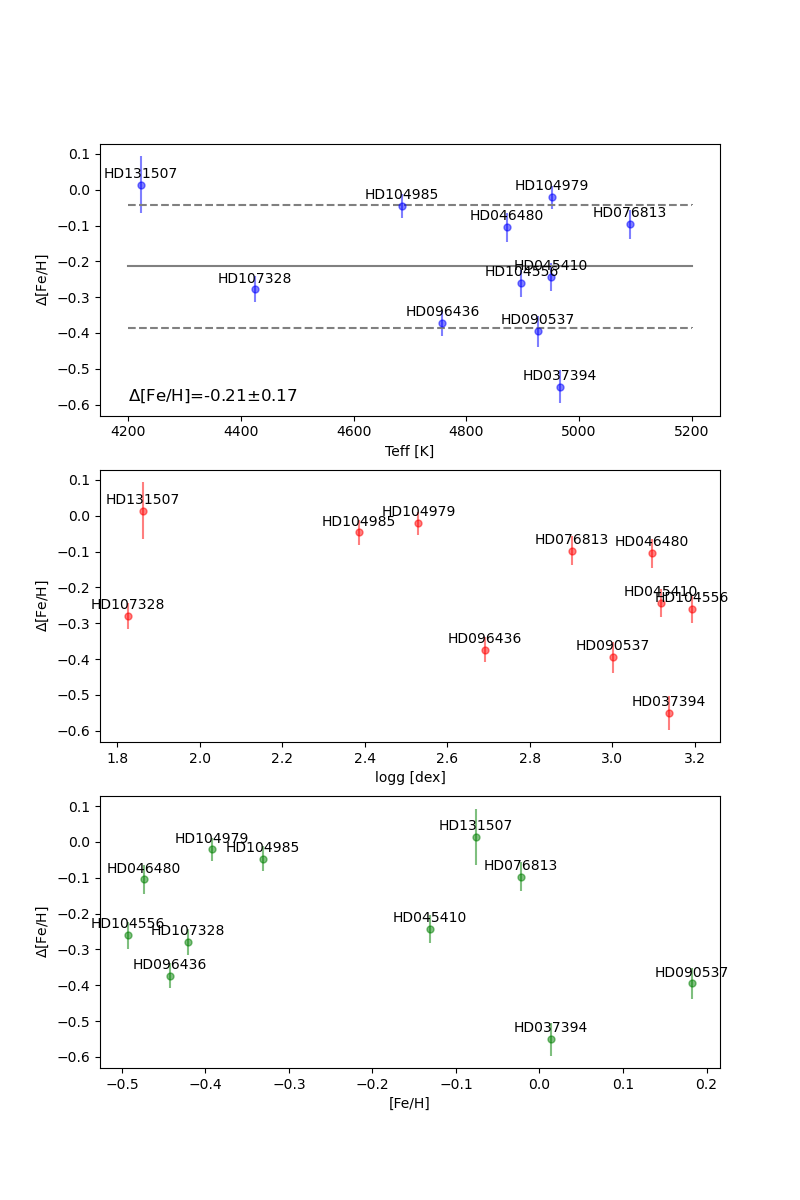}
    \includegraphics[width=0.45\textwidth, trim={0 2cm 1cm 3cm}, clip]{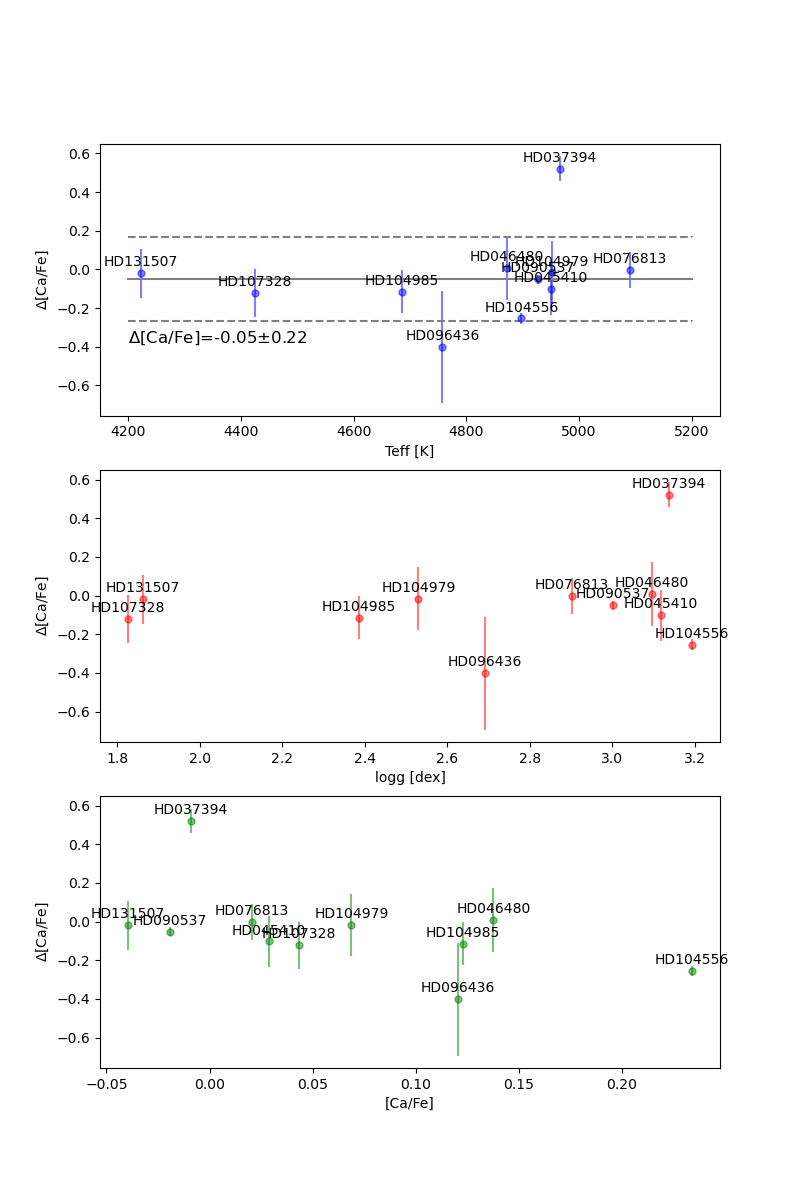}
    \includegraphics[width=0.45\textwidth, trim={0 2cm 1cm 3cm}, clip]{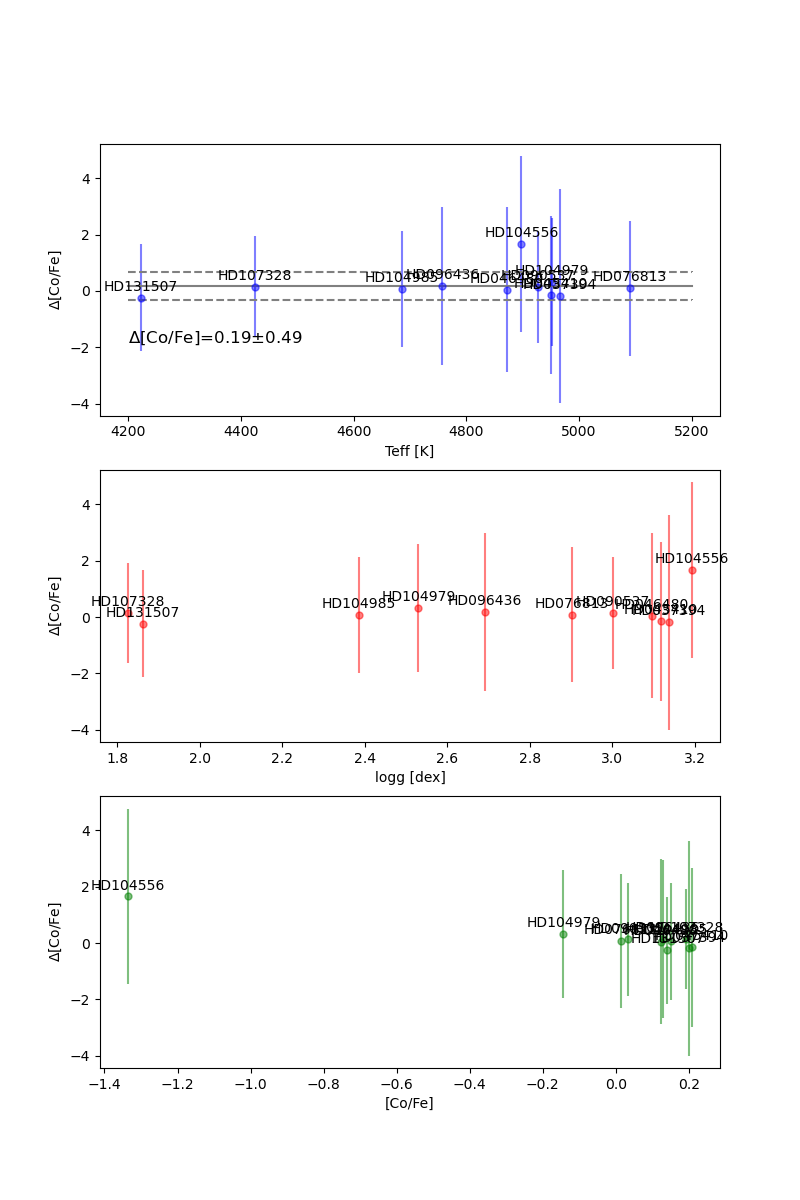}
    \includegraphics[width=0.45\textwidth, trim={0 2cm 1cm 3cm}, clip]{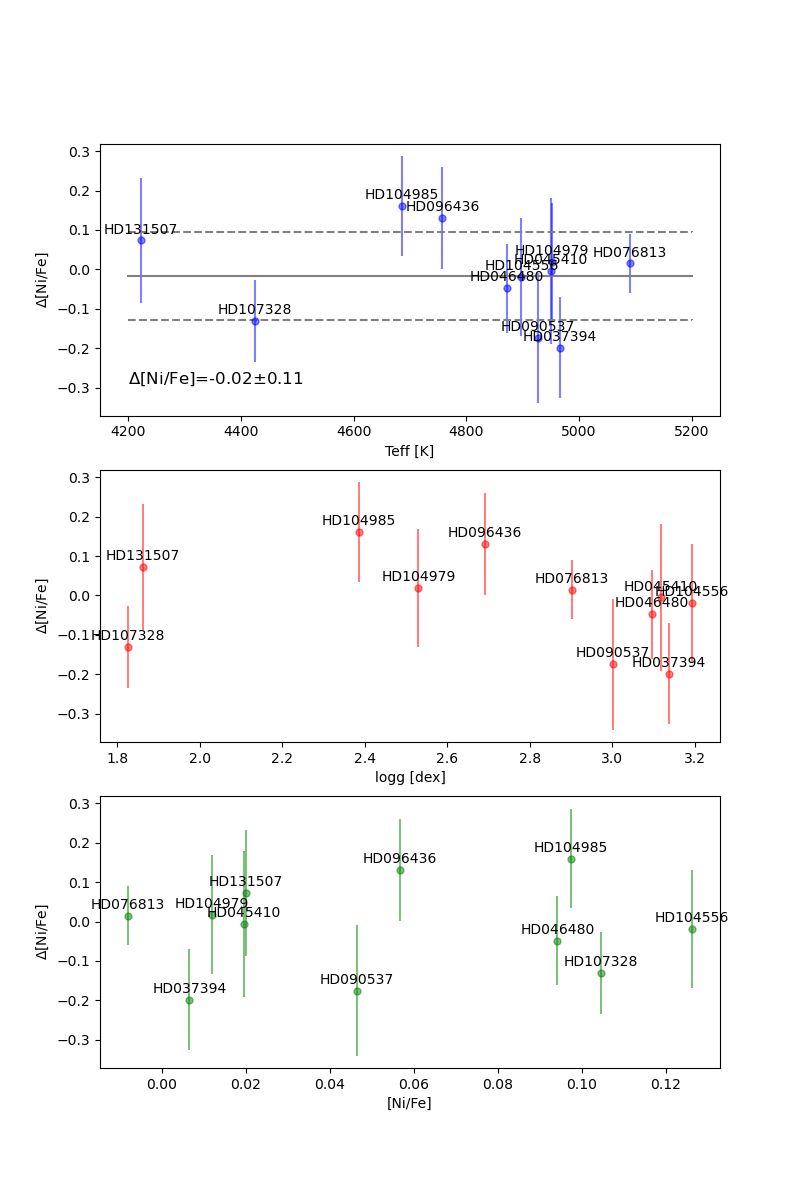}
  \caption{Difference between the chemical abundances obtained from MEGASTAR and APOGEE spectra, with respect to \teff, \logg and abundance (in the sense APOGEE$-$MEGASTAR). Top left panels: [Fe/H]. Top right panels: [Ca/Fe]. Bottom left panels: [Co/Fe]. Bottom right panels: [Ni/Fe].}
  \label{CaFe}
\end{figure*}

Figures \ref{fig:A11} through \ref{fig:Be1023} show zoom-in plots into the spectra of the six red-giant stars in Auner 1 and Berkeley 102 used for abundance analysis. Figure \ref{CaFe} shows the results of our differential analysis of APOGEE and MEGASTAR spectra.

\begin{table*}
\begin{center}
\caption{List of lines for red giants observed with the HR-R grating. The gfflag and synflag columns are taken from \citet{heiter2021}.}
\vspace{-0.5cm} 
\begin{multicols}{2} 
\setlength{\tabcolsep}{1mm}
\begin{tabular}{llll}
    \hline
    \multicolumn{1}{c}{Wavelength} & \multicolumn{1}{c}{Element}& \multicolumn{1}{c}{gfflag}&\multicolumn{1}{c}{synflag}  \\
    \multicolumn{1}{c}{[nm]} &\\
    \hline
    640.80178 & Fe 1 & Y & U \\
    641.16481 & Fe 1 & Y & Y \\
    641.99419 & Fe 1 & - & - \\
    642.13652 & Fe 1 & Y & N \\
    643.08439 & Fe 1 & Y & Y \\
    643.26761 & Fe 2 & Y & Y \\
    643.90687 & Ca 1 & Y & U \\
    645.638   & Fe 2 & U & Y \\
    646.9168  & Fe 1 & U & N \\
    647.5624  & Fe 1 & Y & Y \\
    648.18711 & Fe 1 & Y & Y \\
    648.27987 & Ni 1 & U & Y \\
    648.39434 & Fe 1 & N & Y \\
    649.37822 & Ca 1 & Y & Y \\
    649.49786 & Fe 1 & Y & Y \\
    649.65042 & Fe 1 & U & U \\
    649.68851 & Ba 2 & Y & U \\
    649.89337 & Fe 1 & Y & U \\
    651.60756 & Fe 2 & Y & U \\
    651.83657 & Fe 1 & Y & U \\
    653.2873  & Ni 1 & Y & Y \\
    654.62381 & Fe 1 & Y & Y \\
    655.4223  & Ti 1 & Y & Y \\
    655.6062  & Ti 1 & Y & U \\ 
    656.92173 & Fe 1 & U & U \\
    657.42272 & Fe 1 & Y & Y \\
    \hline
\end{tabular}

\begin{tabular}{llll}
    \hline
    \multicolumn{1}{c}{Wavelength} & \multicolumn{1}{c}{Element} & \multicolumn{1}{c}{gfflag}&\multicolumn{1}{c}{synflag}\\
    \multicolumn{1}{c}{[nm]} &\\
    \hline
    657.50161 & Fe 1 & Y & - \\
    658.11981 & Fe 1 & Y & U \\
    658.63117 & Ni 1 & Y & Y \\
    659.28914 & Fe 1 & Y & U \\
    659.38686 & Fe 1 & Y & U \\
    659.75578 & Fe 1 & U & Y \\
    660.80194 & Fe 1 & U & U \\
    660.91017 & Fe 1 & Y & U \\
    660.96749 & Fe 1 & - & - \\
    661.37975 & Fe 1 & - & - \\
    662.49914 & Fe 1 & Y & N \\
    663.37535 & Fe 1 & Y & U \\
    664.363   & Ni 1 & Y & Y \\
    664.50925 & Eu 2 & Y & U \\
    666.33937 & Fe 1 & Y & - \\
    667.79772 & Fe 1 & Y & - \\
    670.35663 & Fe 1 & U & Y \\
    670.51014 & Fe 1 & Y & Y \\
    671.03181 & Fe 1 & N & Y \\
    672.6666  & Fe 1 & N & Y \\
    673.315   & Fe 1 & Y & Y \\
    673.95202 & Fe 1 & Y & Y \\
    675.0151  & Fe 1 & Y & Y \\
    676.7772  & Ni 1 & Y & Y \\
    677.10345 & Co 1 & Y & Y \\
    677.23133 & Ni 1 & N & Y \\
    \hline
\end{tabular}
\end{multicols}
\label{tab:line_list_MEGARA}
\end{center}
\end{table*}

\begin{table*}
\setlength{\tabcolsep}{1mm}
\begin{center}
\caption{Orbital parameters of the two clusters obtained with different Galactic potentials.}
\begin{tabular}{llllllll}
\hline
 Cluster& Potential&  $R_{\rm apo}$ [kpc] &          $R_{\rm peri}$ [kpc] &        $R_{\rm birth}$ [kpc] &             $R_{\rm mean}$ [kpc] &            e &         $Z_{\rm max}$ [kpc]\\
\hline
Auner~1& MW2014& 16.0 $\pm$ 0.5 &  12.0 $\pm$ 0.9 &  14.1 $\pm$ 1.5 &  14.0 $\pm$ 1.0 &  0.14 $\pm$ 0.02 &   1.0 $\pm$ 0.1 \\
Berkeley~102& MW2014 & 19.2 $\pm$ 1.7 &  15.0 $\pm$ 0.6 &  17.0 $\pm$ 1.8 &  17.0 $\pm$ 1.8 &  0.12 $\pm$ 0.03 &  1.7 $\pm$ 0.3 \\
Auner~1& MW2014+Bar& 15.8 $\pm$ 0.5 &  11.2 $\pm$ 1.0 &  13.6 $\pm$ 1.8 &   13.5 $\pm$ 1.1 &  0.17 $\pm$ 0.03 &  1.0 $\pm$ 0.1 \\
Berkeley~102& MW2014+Bar& 18.4 $\pm$ 1.7 &  14.8 $\pm$ 0.6 &  16.7 $\pm$ 1.7 &  16.6 $\pm$ 1.8 &  0.11 $\pm$ 0.03 &   1.6 $\pm$ 0.3 \\
Auner~1& MW2014+Arms&16.5 $\pm$ 0.7 &  11.5 $\pm$ 0.7 &   14.2 $\pm$ 1.6 &  14.0 $\pm$ 1.0 &  0.18 $\pm$ 0.01 &  1.1 $\pm$ 0.2 \\
Berkeley~102& MW2014+Arms& 21.2 $\pm$ 2.2 &  14.7 $\pm$ 0.7 &  17.7 $\pm$ 2.2 &  18.0 $\pm$ 2.3 &  0.18 $\pm$ 0.04 &  2.4 $\pm$ 0.7 \\
Auner~1& MW2014+Bar+Arms&16.3 $\pm$ 0.6 &  11.1 $\pm$ 0.9 &  14.1 $\pm$ 1.7 &  13.7 $\pm$ 1.0 &  0.19 $\pm$ 0.02 &  1.1 $\pm$ 0.2 \\
Berkeley~102& MW2014+Bar+Arms&19.7 $\pm$ 2.4 &  14.6 $\pm$ 0.6 &  17.2 $\pm$ 2.0 &  17.2 $\pm$ 2.5 &  0.15 $\pm$ 0.05 &  2.2 $\pm$ 0.7 \\

\hline
\end{tabular}
\label{tab:Orbitas}
\end{center}
\end{table*}

\begin{table}
\setlength{\tabcolsep}{1mm}
\begin{center}
\caption{Solar abundances obtained in this paper from our MEGARA spectra, compared with \citet[][GAS07]{Grevesse2007} and \citet[][AGS09]{Asplund2009}.}
\begin{tabular}{lcccc}
\hline
Element& This work & GAS07 & AGS09 \\
\hline
\mbox{Fe\,{\sc i}}& 7.43$\pm$0.01 & 7.45$\pm$0.05&  7.50 $\pm$0.04 \\
\mbox{Ca\,{\sc i}}& 6.16$\pm$0.02  & 6.31$\pm$0.04 &  6.34 $\pm$0.04  \\
\mbox{Ti\,{\sc i}}& 4.89$\pm$0.01  & 4.90$\pm$0.06 &   4.95$\pm$0.05  \\
\mbox{Co\,{\sc i}}& 4.81$\pm$0.04 & 4.92$\pm$0.08 &  4.99 $\pm$0.07 \\
\mbox{Ni\,{\sc i}}& 6.19$\pm$0.03 & 6.23$\pm$0.04 &  6.22 $\pm$0.04 \\
\mbox{Ba\,{\sc ii}}& 2.20$\pm$0.05 & 2.17$\pm$0.07 &  2.18 $\pm$0.09 \\       
\mbox{Eu\,{\sc ii}}& 0.67$\pm$0.06 & 0.52$\pm$0.06 &  0.52$\pm$0.04 \\
        \hline
\end{tabular}
\label{tab:solar_MEGARA}
\end{center}
\end{table}

\clearpage

\end{appendix}

\end{document}